\documentclass[utf8]{frontiersFPHY}
\usepackage{physics}
\usepackage{url,hyperref,lineno,microtype}
\usepackage{subcaption}
\usepackage[onehalfspacing]{setspace}

\newcommand{\heff}{$H_{\rm eff}$}
\newcommand{\heffs}{$H_{\rm eff}$s}
\newcommand{\qbox}{$\hat{Q}$-box}
\newcommand{\vlwk}{$V_{{\rm low}\mbox{-}k}$}
\newcommand{\vlwks}{$V_{{\rm low}\mbox{-}k}$s}
\newcommand{\zbb}{$0\nu\beta\beta$}
\newcommand{\dbb}{$2\nu\beta\beta$}
\newcommand{\nme}{$M^{0\nu}$}
\newcommand{\nmes}{$M^{0\nu}$s}
\newcommand{\nmed}{$M^{2\nu}_{\rm GT}$}
\newcommand{\nmeds}{$M^{2\nu}_{\rm GT}$s}

\setcitestyle{square} 
\def\keyFont{\fontsize{8}{11}\helveticabold }
\def\firstAuthorLast{L. Coraggio and N. Itaco} 
\def\Authors{Luigi Coraggio\,$^{1,*}$ and Nunzio Itaco\,$^{1,2,*}$}
\def\Address{$^{1}$Istituto Nazionale di Fisica Nucleare, Sezione di Napoli, Napoli, Italy\\
  $^{2}$Dipartimento di Matematica e Fisica, Universit\`a degli
  Studi della Campania ``Luigi Vanvitelli'', Caserta, Italy}

\def\corrAuthor{Luigi Coraggio \\ Sezione INFN di Napoli, Complesso Universitario di Monte  Sant'Angelo, Via Cintia - I-80126 Napoli, Italy}
\def\corrEmail{luigi.coraggio@na.infn.it}

\begin{document}
\onecolumn
\firstpage{1}

\title[Perturbative Approach to Effective Shell-Model
Hamiltonians and Operators]{Perturbative Approach to Effective Shell-Model
  Hamiltonians and Operators} 

\author[\firstAuthorLast ]{\Authors} 
\address{} 
\correspondance{} 

\extraAuth{Nunzio Itaco \\ Dipartimento di Matematica e Fisica, Universit\`a degli
  Studi della Campania ``Luigi Vanvitelli'', Viale A. Lincoln 5, Caserta, I-81100, Italy \\ nunzio.itaco@unicampania.it}

\maketitle

\begin{abstract}

\section{}
The aim of this work is to present an overview of the derivation of
the effective shell-model Hamiltonian and decay operators within
many-body perturbation theory, and to show the results of selected
shell-model studies based on their utilisation.
More precisely, we report some technical details that are needed by
non-experts to approach the derivation of shell-model Hamiltonians and
operators starting from realistic nuclear potentials, in order to
provide some guidance to shell-model calculations where the
single-particle energies, two-body matrix elements of the residual
interaction, effective charges and decay matrix elements, are all
obtained without resorting to empirical adjustments.
On the above grounds, we will present results of studies
  of double-$\beta$ decay of heavy-mass nuclei where shell-model
  ingredients are derived from theory, so to
assess the reliability of such a way to shell-model investigations.
Attention will be also focussed on the relevant aspects
that are connected to the behavior of the perturbative expansion,
whose knowledge is needed to establish limits and perspectives of this
approach to nuclear structure calculations.

\tiny
\keyFont{ \section{Keywords:} Nuclear shell model, effective
  interactions, many-body perturbation theory, nuclear forces}

\end{abstract}

\section{Introduction}
\label{intro}
The present paper is devoted to the presentation of the formal details
of the derivation of effective shell-model Hamiltonians (\heff) and
decay operators by way of a perturbative approach, and to review a
large sample of its most recent applications to the study of
spectroscopic properties of atomic nuclei.
The goal of our work is to provide a useful tool for those
practitioners who are interested in employing shell-model
single-particle energies, two-body matrix elements, effective charges,
magnetic-dipole and $\beta$-decay operators, which are produced by way
of many-body theory, without resorting to parameters that are
empirically adjusted to reproduce a selection of observables.

As is well known, the nuclear shell model (SM) is widely considered the basic
theoretical tool for the microscopic description of nuclear structure
properties.
Nuclear shell model is based on the ansatz that each nucleon inside
the nucleus moves independently from the others, in a spherically
symmetric mean field plus a strong spin-orbit term.
This first-approximation depiction of a nucleus is supported by the
observation of ``magic numbers'' of protons and/or neutrons,
corresponding to nuclei which are more tightly bound than their
neighbors.

These considerations lead to depict the nucleons as arranging
themselves into groups of energy levels, the ``shells'', well
separated from each other.
The main product of the shell-model scheme is the reduction of the
complex nuclear many-body problem to a very simplified one, where only
a few valence nucleons interact in a reduced model space spanned by a
single major shell above an inert core.

The cost that has to be paid for such a simplification is that
shell-model wave functions, describing the independent motion of
individual nucleons, do not include the correlations which are induced
by the strong short-range bare interaction, and therefore could be
very different from the real wave functions of the nuclei.
The shell-model Hamiltonian, which will be introduced in the following
section, contains one- and two-body components whose characterizing
parameters, namely the single-particle (SP) energies and two-body
matrix elements (TBMEs) of the residual interaction, account for the
degrees of freedom that are not explicitly included in the truncated
Hilbert space of the configurations.
As a matter of fact, SP energies and TBMEs should be determined to
include, in an effective way, the excitations both of core nucleons
and of the valence nucleons into the shells above the model space.

The way to the effective SM Hamiltonian may follow two distinct paths.

One approach is phenomenological, that is the one- and two-body
components of the Hamiltonian are adjusted to reproduce a selected set
of experimental data.
This can be done either using an analytical expression for the
residual interaction with adjustable parameters, or treating the
Hamiltonian matrix elements directly as free parameters (see
\citep{Elliott69a,Talmi03}).

This has been, during seventy years and more of SM
calculations, a very successful tool to reproduce a huge amount of
data and to describe some of the most fundamental physical properties
of the structure of atomic nuclei.
In this regard, it is worth to mention the review by Caurier {\it et
  al.} \citep{Caurier05} for an interesting discussion about the
properties of the effective SM Hamiltonian; a few more references and
discussion will be reported in the following section.

The alternative way to the construction of \heff~ is to start from
realistic nuclear forces - two- and three-body potentials (if
possible) - and derive the effective Hamiltonian in the framework of
the many-body theory, namely a \heff~ whose eigenvalues belong
to the set of eigenvalues of the full nuclear Hamiltonian, defined in
the whole Hilbert space.

To this end, we need a similarity transformation which arranges,
within the full Hilbert space of the configurations, a decoupling of
the model space $P$ where the valence nucleons are constrained from
its complement $Q=1-P$.

Nowadays, this may achieved within the framework of the {\it ab
  initio} methods, which aim to solve the full Hamiltonian of $A$
nucleons by employing controlled truncations of the accessible degrees
of freedom.
However, this approach is strictly constrained by the advance in
computational power, and, even if successful, is currently confined to few
nuclear mass regions.
A comprehensive report of possible ways to tackle the problem of the
derivation of \heff~ starting from {\it ab initio} methods can be found
in Ref. \citep{Stroberg19}, where the authors review also some SM
applications and results.

Our work will be focused on the perturbative expansion of the
effective SM Hamiltonian, that is grounded in the
  energy-independent linked-diagram perturbation theory \citep{Kuo90},
  which has been extensively used in shell-model calculations during
  the last fifty years (see also review papers \citep{Hjorth95,Coraggio09a}).

The earlier attempt along this line has been made by Bertsch
\citep{Bertsch65}, who employed as interaction vertices the matrix
elements of the reaction matrix $G$ derived from the Kallio-Kolltveit
potential \citep{Kallio64} to study the role played by the
core-polarization diagram at second order in perturbation theory,
accounting for one-particle-one-hole ($1p-1h$) excitations above the
Fermi level of the core nucleons.
The results of this work evidenced that the contribution of such a
diagram to \heff~ was about $30\%$ of the first-order two-body matrix
element, when considering the open-shell nuclei $^{18}$O and
$^{42}$Sc outside doubly-closed cores $^{16}$O and $^{40}$Ca,
respectively.  

Then, it came the seminal paper by Tom Kuo and  Gerry Brown
\citep{Kuo66}, which is a true turning point in nuclear
structure theory.
It has indeed been the first successful attempt to perform a
shell-model calculation starting from the free
nucleon-nucleon ($NN$) Hamada-Johnston potential (HJ)
\citep{Hamada62}, and resulted in a quantitative description of the
spectroscopic properties of $sd$-shell nuclei.

The TBMEs of the $sd$-shell effective interaction in Ref. \citep{Kuo66}
were derived starting from the HJ potential, the hard-core component
being renormalized via the calculation of the reaction-matrix $G$.
The matrix elements of $G$ were then employed as interaction vertices
of the perturbative expansion of \heff, including terms up to second
order in $G$.

The TBMEs obtained within this approach were used to calculate the
energy spectra of $^{18}$O and $^{18}$F and provided good agreement
with experiment.
Moreover, these matrix elements, as well as those derived two years
later for SM calculations in the $fp$ shell \citep{Kuo68a}, have
become the backbone of the fine tuning of successful empirical
\heffs~such as the USD \citep{Brown88} and the KB3G potentials
\citep{Caurier05,Poves01}.

The theoretical framework has evolved between the end of 1960s and
beginning of 1970s, thanks to the introduction of the folded-diagrams
expansion which has formally defined the correct procedure for the
perturbative expansion of effective shell-model Hamiltonians
\citep{Brandow67,Kuo71}.

In the forthcoming sections we are going to present in detail the
derivation of \heff~ and of consistent effective SM decay operators,
according to the theoretical framework of the many-body perturbation
theory. 
The core of our approach is the perturbative expansion of two vertex
functions, the so-called \qbox~and $\hat{\Theta}$-box, in
  terms of irreducible valence-linked Goldstone diagrams.
The \qbox~is then employed to solve non-linear matrix equations to
obtain \heff~ by way of iterative techniques \citep{Suzuki80}, while
the latter together with the $\hat{\Theta}$-box are the main
ingredients to derive the effective decay operators \citep{Suzuki95}.

Our paper is organized as follows.

In the next section we will present a general overview of the SM
eigenvalue problem, and of the derivation of the effective SM
Hamiltonian.

In section \ref{perturbative} we will tackle the problem on the
grounds of the Lee-Suzuki similarity transformation
\citep{Suzuki80,Lee80}, and in its sections we will introduce the
iterative procedures to solve the decoupling equation which provide
this similarity transformation into \heff~, both for degenerate and
non-degenerate model spaces.
Two sections will be devoted also to the perturbative expansion of
the \qbox~vertex function and to the derivation of effective SM decay
operators.

In section \ref{applications} we will show the results of
investigations about the double-$\beta$ decay of $^{130}$Te and
$^{136}$Xe, and discuss the perturbative properties of \heff~and
effective shell-model decay operators.

In the last section, a summary of our present work will be reported.

\section{General Overview}
\label{overview}
As mentioned in the Introduction, the shell model, introduced into
nuclear physics seventy years ago \citep{Mayer49,Haxel49}, is based on
the assumption that, as a first approximation, each nucleon (proton or
neutron) inside the nucleus moves independently in a spherically
symmetric potential representing the average interaction with the
other nucleons.
This potential is usually described by a Woods-Saxon or harmonic
oscillator (HO) potential including a strong spin-orbit term.
The inclusion of the latter term is crucial providing single-particle
states clustered in groups of orbits lying close in energy (shells). 
Each shell is well separated in energy from the others, and this
enables to schematize the nucleus as an inert core, made up 
by shells filled up with neutrons and protons paired to a total
angular momentum $J=0^+$, plus a certain number of external nucleons,
the so-called ``valence" nucleons.
This extreme single-particle shell model is able to
  successfully describe various nuclear properties
  \citep{Mayer55}, as, for instance, the angular momentum and parity of the
  ground-states in odd-mass nuclei. 
However, it is clear that to describe the low-energy structure of
nuclei with two or more valence nucleons the ``residual" interaction between 
the valence nucleons has to be considered explicitly, the term
 residual meaning that part of the interaction which is not taken into
 account by the central potential.
The inclusion of the residual interaction removes the degeneracy of
the states belonging to the same configuration and produces a mixing
of different configurations. 

Let us now use the simple nucleus $^{18}$O to introduce some common
terminologies used in effective interaction theories. 

Suppose we want to calculate the properties of the low-lying states in $^{18}$O.
Then, we must solve the Schr{\"o}dinger equation
\begin{equation}
H | \Psi_{\nu} \rangle  = E_{\nu} | \Psi_{\nu} \rangle \label{eq1},
\end{equation}

\noindent
where
\begin{equation}
 H=H_0+H_1  \label{defh},
\end{equation}
and 
\begin{equation}
H_0= \sum_{i=1}^A \left(\frac{p_i^2}{2m}+U_i \right)  \label{defh0},
\end{equation}
\begin{equation}
H_1=\sum_{i<j=1}^{A} V^{NN}_{ij}-\sum_{i=1}^AU_i \label{defh1}~. 
\end{equation}
An  auxiliary one-body potential $U_i$ has been introduced in order to
break up the nuclear Hamiltonian as the sum of a one-body term
$H_0$, which describes the independent motion of the nucleons, and 
the residual interaction $H_1$.
It is worth pointing out that in the following, for sake of
simplicity and without any loss of generality, we will assume that the
interaction between the nucleons is described by a two-body force
only, neglecting three-body contributions.
The generalization of the formalism to include 3N forces may be found
in Refs. \citep{Fukui18,Ma19}.

It is customary to choose an auxiliary one-body potential $U$ of
convenient mathematical form, e.g. the harmonic oscillator potential
\begin{equation}
U=\sum_{i=1}^A\frac{1}{2}m\omega r_i^2 \label{defu}~. 
\end{equation}

\begin{figure}[h!]
\begin{center}
\includegraphics[width=6cm]{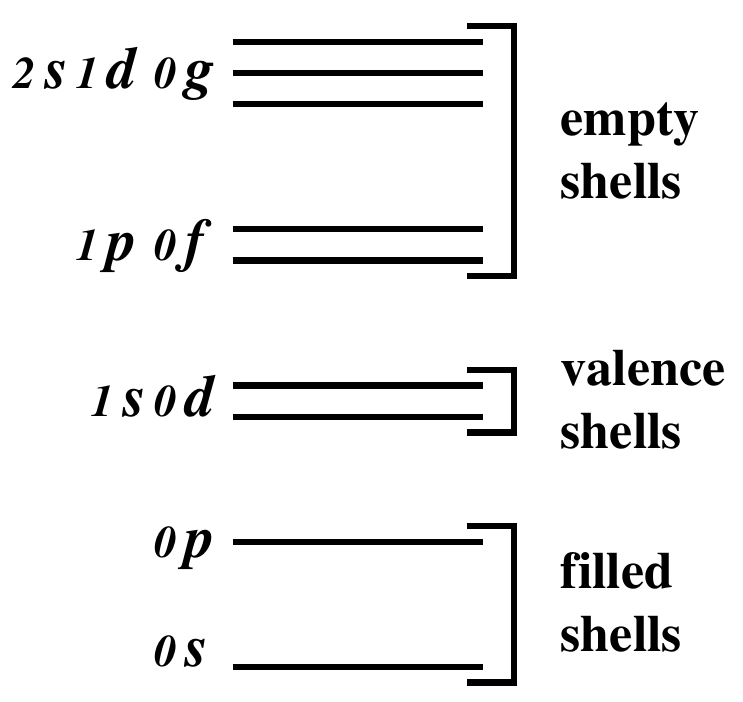}
\end{center}
\caption{Energy shells which characterize the core, valence space, and
empty orbitals for $^{18}$O}\label{H0-spectrum}
\end{figure}

In Fig. \ref{H0-spectrum} we report the portion of the $H_0$ spectrum
relevant for $^{18}$O.

We expect that the wave functions of the low-lying states in $^{18}$O
are dominated by components with a closed $^{16}$O core (i.e. the $0s$
and $0p$ orbits are filled) and two neutrons in the valence orbits
$1s$ and $0d$.

Thus, we choose a model space which is spanned by the vectors

\begin{equation}
|\Phi_i \rangle = \sum_{\alpha \beta \in~ valence~space} C^i_{\alpha \beta}[ a^{\dagger}_\alpha a^{\dagger}_\beta ]_i
| c \rangle~,~~~ i=1,...,d, \label{eigenfunc}
\end{equation}

\noindent
where $|c \rangle$ represents the unperturbed $^{16}$O core, as
obtained by completely filling the $0s$ and $0p$ orbits

\begin{equation}
| c \rangle = \prod_{\alpha \in~ filled~ shells} a^{\dagger}_\alpha | 0 \rangle~,
\end{equation}

\noindent
and the index $i$ stands for all the other quantum numbers needed to
specify the state (e.g. the total angular momentum).

\begin{figure}[h!]
\begin{center}
\includegraphics[width=10cm]{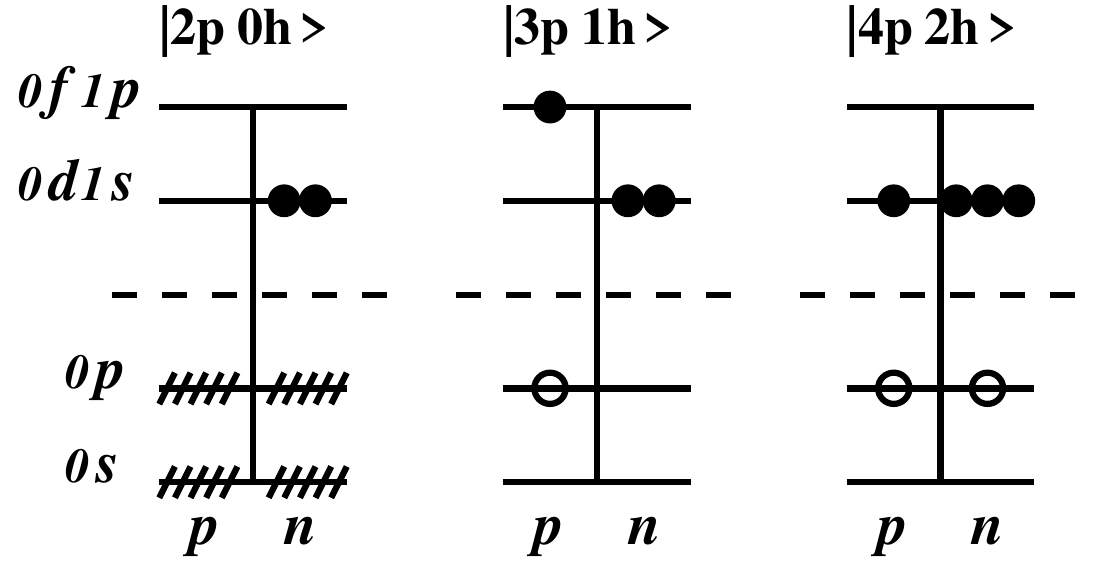}
\caption{Some $^{18}$O shell-model configurations}\label{18Omsp}
\end{center}
\end{figure}

To sketch pictorially the situation, we report in Fig. \ref{18Omsp}
some SM configurations labeled in terms of particles and
holes with respect to the $^{16}$O core.

\begin{figure}[h!]
\begin{center}
\includegraphics[width=11cm]{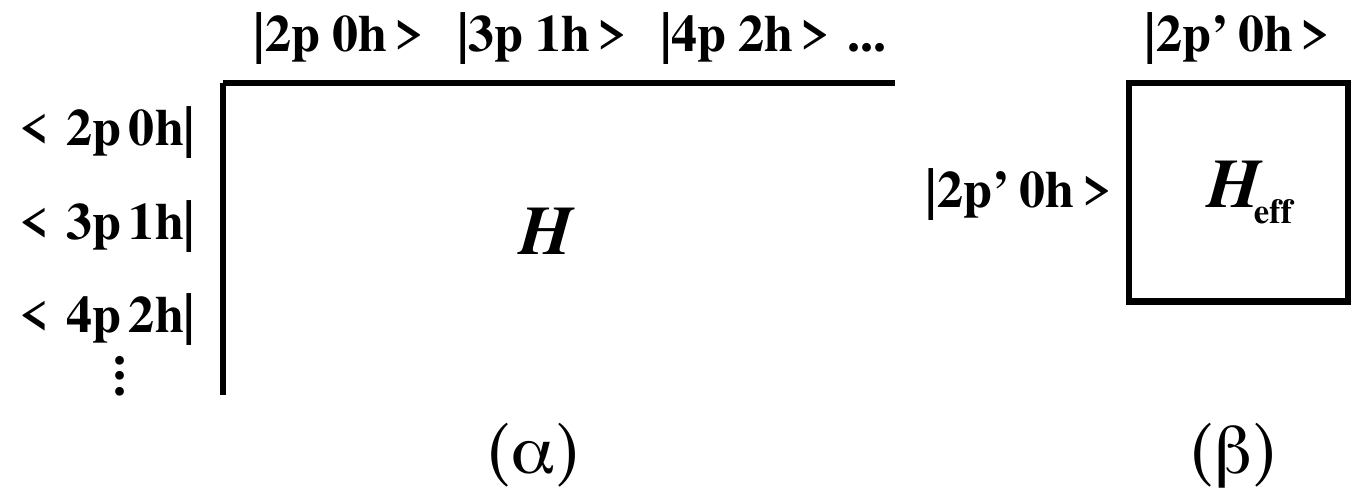}
\caption{Representation of the matrices $H$ and $H_{\rm eff}$ for
  $^{18}$O}\label{18Omatrices}
\end{center}
\end{figure}

To solve Eq. (\ref{eq1}) using basis vectors like those shown in
Fig. \ref{18Omsp} amounts to diagonalizing the infinite matrix $H$ in
Fig. \ref{18Omatrices}.
This is unfeasible, so we want to reduce this huge matrix to a smaller
one, \heff, requiring that the eigenvalues of the latter belong to the
set of the eigenvalues of the former.
The notation $|{\rm 2p'~0h} \rangle$ represents a configuration with a
closed $^{16}$O core plus 2 particles constrained interact in the $sd$
shell.

More formally, it is convenient to introduce the projection operators
$P$ and $Q=1-P$ that project from the complete Hilbert space onto the
model space and its complementary space (excluded space), respectively.

$P$ can be expressed in terms of the vectors in Eq. (\ref{eigenfunc}) as
follows
\begin{equation}
P= \sum_{i=1}^d | \Phi_i \rangle \langle \Phi_i |~, \label{Peq1}
\end{equation}
The projection operators $P$ and $Q$
satisfy the properties
\begin{equation}
P^2=P, ~~ Q^2=Q, ~~ PQ=QP=0 ~. \label{Peq2}
\end{equation}

The scope of the effective SM interaction theory is to
  transform the eigenvalue problem of Eq. (\ref{eq1}) into a reduced
  model-space eigenvalue problem
\begin{equation}
PH_{\rm eff}P | \Psi_{\alpha} \rangle  = (E_{\alpha}-E_C) P | \Psi_{\alpha}
\rangle \label{eq3}~,
\end{equation}

where $E_C$ is the true energy of the core; i.e. in the present case, the
true ground-state energy of $^{16}$O. 

As mentioned in the Introduction, there are two main lines of 
attack to derive \heff :
\begin{itemize}
  \item using a phenomenological approach
  \item starting from the bare nuclear interactions by means of a
    well-suited many-body theory.
\end{itemize}

In the phenomenological approach, empirical effective interactions
containing adjustable parameters are introduced and modified to fit a
certain set of experimental data or the two body-matrix elements
themselves are treated as free parameters.
This approach is very successful and we refer to several excellent
reviews \citep{Talmi03,Caurier05,ABrown01,Brown19,Otsuka20} for a
complete discussion on the topic.

Nowadays, there are several approaches to derive an effective SM
Hamiltonian starting from the bare interaction acting among nucleons.
As a matter of fact, aside the well-established approaches based
on the many-body perturbation theory \citep{Kuo90} or on
  the Lee-Suzuki transformation \citep{Suzuki80,Lee80}, novel
non-perturbative methods like valence-space in-medium SRG (VS-IMSRG)
\citep{Morris15}, shell-model coupled cluster (SMCC) \citep{Sun18}, or
no-core shell-model (NCSM) with a core
\citep{Lisetskiy08,Lisetskiy09,Dikmen15,Smirnova19} based on the
Lee-Suzuki similarity transformation, are now available.
These non-perturbative approaches are firmly rooted in many-body
theory and provide somehow different paths to \heff.
They can be derived in the same general theoretical framework
expressing \heff~as the result of a similarity transformation acting
on the original Hamiltonian 
\begin{equation}
    H_{\mathrm{eff}} = e^{{\ensuremath{\mathcal{G}}}}He^{-{\ensuremath{\mathcal{G}}}},
\end{equation}
\noindent
where the transformation is parametrized as the exponential of a
generator ${\ensuremath{\mathcal{G}}}$, and is such that the
decoupling condition is satisfied
\begin{equation}\label{decouple}
    QH_{\mathrm{eff}}P =0.
\end{equation}

In Ref. \citep{Stroberg19}, it can be found a very detailed discussion
showing how the different methods (perturbative and non-perturbative)
can be derived in such a general framework and describing the
corresponding approximation schemes employed in each approach.

As written in the Introduction, the aim of present review is to
describe in detail the perturbative approach to the derivation of
\heff, topic that will be discussed in the next section.
We refer to the already cited review paper by Stroberg {\it et
  al} \citep{Stroberg19} for an exhaustive description of the
alternative methods.

\section{The perturbative expansion of effective shell-model
  operators}\label{perturbative}
\vspace{0.2truecm}
\subsection{The Lee-Suzuki similarity transformation} \label{sim}
\vspace{0.1truecm}
Here, we will present the formalism of the derivation of the effective
SM Hamiltonian, according to the similarity transformation introduced
by Lee and Suzuki \citep{Lee80}.
It is worth noting that this approach has been very successful since it
is amenable of a straightforward perturbative expansion of \heff~for
open-shell systems outside a closed core, whereas in other approaches -
such as, for example, the oscillator based effective theory (HOBET)
proposed by Haxton and Song \cite{Haxton00} or the coupled-cluster
similarity transformation \cite{Kummel78} - the factorization of the core
configurations with respect to the valence nucleons is far more
complicated to perform.

We start from the Schr\"odinger equation for the $A$-nucleon system,
defined in the whole Hilbert space:
\begin{equation}
H | \Psi_{\nu} \rangle  = E_{\nu} | \Psi_{\nu} \rangle \label{Eq1}~.
\end{equation}

As already mentioned, within the SM framework an auxiliary one-body
potential $U$ is introduced to express the nuclear
  Hamiltonian as the sum of an unperturbed one-body mean-field term
  $H_0$, plus the residual interaction Hamiltonian $H_1$. 
The full Hamiltonian $H$ is then rewritten in terms of $H_0,H_1$, as
reported in Eqs. (\ref{defh}-\ref{defh1}).

According to the nuclear SM that we have introduced in the previous
section, the nucleus may be depicted as a frozen core, composed by a
number of nucleons which fill a certain number of energy shells
generated by the spectrum of the one-body Hamiltonian $H_0$, plus a
remainder of $n$ interacting valence nucleons moving in the mean field
$H_0$.
 
The large energy gap between the shells allows to consider the $A-n$
core nucleons, filling completely the shells which are the lowest in
energy, as inert.
The SP states accessible to the valence nucleons are those belonging
to the major shell just placed (in energy) above the closed core.
The configurations allowed by the valence nucleons within this major
shell define a reduced Hilbert space, the model space,  in terms of a
finite subset of $d$ eigenvectors of $H_0$, as expressed in
Eq. (\ref{eigenfunc}).

We then consider the projection operators $P$ (see Eq.(\ref{Peq1}))
and $Q=1-P$, which project from the complete Hilbert space onto the
model space and its complementary space, respectively, and satisfy the
properties in Eq. (\ref{Peq2}).

The goal of a SM calculation is to reduce the eigenvalue problem of
Eq. (\ref{Eq1}) to the model-space eigenvalue problem
\begin{equation}
H_{\rm eff} P | \Psi_{\alpha} \rangle  = E_{\alpha} P | \Psi_{\alpha}
\rangle \label{Eq3}~,
\end{equation}

\noindent
where $\alpha=1,..,d$ and $H_{\rm eff}$ is defined only in the model
space.

This means that we are looking for a new Hamiltonian $\mathcal{H}$
whose eigenvalues are the same of the Hamiltonian $H$ for the
$A$-nucleon system, but satisfies the decoupling equation between the
model space $P$ and its complement $Q$:

\begin{equation}
Q \mathcal{H} P=0 ~, \label{deceq1}
\end{equation}

\noindent
which guarantees that the desired effective Hamiltonian is $H_{\rm eff}= P
\mathcal{H} P$.

The Hamiltonian $\mathcal{H}$ should be obtained by way of a
similarity transformation defined in the whole Hilbert space:

\begin{equation}
\mathcal{H}=X^{-1} H X ~.
\end{equation}
\noindent

Of course, the class of transformation operators $X$ that satisfy the
decoupling equation (\ref{deceq1}) is infinite, and Lee and Suzuki
\citep{Suzuki80,Lee80} have proposed an operator $X$ defined as
$X=e^{\omega}$.
Without loss of generality, $\omega$ can be chosen to satisfy the
following properties:

\begin{equation}
\omega= Q \omega P ~, 
\label{omegapro1}
\end{equation}
\begin{equation}
P \omega P= Q \omega Q = P \omega Q =0 ~. 
\label{omegapro2}
\end{equation}

\noindent
Eq. (\ref{omegapro1}) implies that 

\begin{equation}
\omega ^2 = \omega^3 = ~...~=0 ~. 
\label{omegapro3}
\end{equation}

According to the above equation, $X$ may be written as $X=1+ \omega$,
and consequently we have the following expression for \heff:

\begin{equation}
H_{\rm eff} = P \mathcal{H} P = PHP +PH Q \omega ~. \label{defheff}
\end{equation}

The operator $\omega$ may be calculated by solving the decoupling equation
(\ref{deceq1}), and the latter may be rewritten as

\begin{equation}
Q H P + Q H Q \omega - \omega P H P - \omega P H Q \omega = 0
~. \label{deceq2} 
\end{equation}

The above matrix equation is non-linear and, once the Hamiltonian $H$
is expressed explicitly in the whole Hilbert space, it can be easily solved.
Actually, this is not an easy task for nuclei with mass $A>2$, and, as
mentioned in the previous section, this approach has been employed
only for light nuclei within the {\it ab-initio} framework.

A successful way to the solution of Eq. (\ref{deceq2}) for SM
calculations is the introduction of a vertex function, the \qbox,
which is suitable of a perturbative expansion.

We proceed now to explicit the \qbox~approach towards the derivation
of \heff, and it is important to point out that in the following we
assume our model space to be degenerate:

\begin{equation}
P H_0 P =\epsilon_0 P~. \label{dege}
\end{equation}

Then, thanks to the decoupling equation (\ref{deceq1}), the effective
Hamiltonian $H^{\rm eff}_1=H_{\rm eff} - P H_0 P$ can be expressed as
a function of $\omega$

\begin{equation}
H^{\rm eff}_1 = P \mathcal{H} P - P H_0 P = P H_1 P + P H_1 Q \omega
~. \label{eqqq}
\end{equation}

The above identity, the decoupling equation (\ref{deceq2}), and the
properties of $H_0$ and $H_1$ allow to define recursively
  the effective Hamiltonian $H^{\rm eff}_1$.

First, since $H_0$ is diagonal, we can write the following identity:

\begin{equation}
QHP= QH_1P + QH_0P = QH_1P~.
\end{equation}

Then, the decoupling equation (\ref{deceq2}) can be rewritten in the
following form:

\begin{equation}
Q H_1 P + Q H Q \omega - \omega (P H_0 P + P H_1 P + P H_1 Q \omega) = 
Q H_1 P + QHQ \omega - \omega ( \epsilon_0 P + H_1^{\rm eff}) = 0
~. \label{deceq3} 
\end{equation}

Using this expression of the decoupling equation, we can write a new
identity for the operator $\omega$:

\begin{equation}
\omega = Q \frac{1}{\epsilon_0 -QHQ} Q H_1 P - Q \frac{1}{\epsilon_0
    -QHQ} \omega H^{\rm eff}_1~. \label{omegaq}
\end{equation}

Finally, we obtain a recursive equation by inserting
Eq. (\ref{omegaq}) into the identity (\ref{eqqq}) which defines
$H^{\rm eff}_1$:

\begin{equation}
H^{\rm eff}_1 (\omega) = P H_1 P + P H_1 Q \frac{1}{\epsilon_0 - Q H Q} Q
  H_1 P - P H_1 Q \frac{1}{\epsilon_0 - Q H Q} \omega H^{\rm eff}_1
  (\omega) ~. \label{eqsemifinal}
\end{equation}

We define now the vertex function $\hat{Q}$-box as follows:

\begin{equation}
\hat{Q} (\epsilon) = P H_1 P + P H_1 Q \frac{1}{\epsilon - Q H Q} Q
H_1 P ~, \label{qbox}
\end{equation}

\noindent
that allows to express the recursive equation (\ref{eqsemifinal}) as

\begin{equation}
H^{\rm eff}_1 (\omega) = \hat{Q}(\epsilon_0) - P H_1 Q \frac{1}{\epsilon_0
  - Q H Q} \omega H^{\rm eff}_1 (\omega) ~. \label{eqfinal}
\end{equation}

As can be seen from both Eqs. (\ref{qbox},\ref{eqfinal}),
configurations belonging to the $Q$ space that are close in energy to
the unperturbed energy of model-space configurations (intruder states)
may provide unstable solutions of Eq. \eqref{eqfinal}.
This is the so-called ``intruder-state problem'' as introduced in
Ref. \citep{Schucan72,Schucan73} by Schucan and Weidenm\"uller.
In the following sections we first show two possible iterative
techniques to solve Eq. \eqref{eqfinal}, as suggested by Lee and
Suzuki\citep{Suzuki80}.
These methods, which are based on the calculation of the \qbox~and its
derivatives, are known as the Krenciglowa-Kuo (KK) and the Lee-Suzuki
(LS) techniques.
In particular, we point out that in Ref. \citep{Suzuki80} the authors
have shown that the Lee-Suzuki iterative procedure is convergent even
when there are some intruder states.

Then, we will present other approaches that generalize the derivation
of \heff, based on the calculation of the \qbox, to unperturbed
Hamiltonians $H_0$ which provide non-degenerate model spaces.
\vspace{0.4truecm}

\vspace{0.4truecm}
\subsubsection{The Krenciglowa-Kuo iterative technique}\label{KK}
\vspace{0.1truecm}
The Krenciglowa-Kuo (KK) iterative technique for solving the recursive
equation (\ref{eqfinal}) traces back to the coupling of
Eqs. (\ref{eqfinal}) and (\ref{omegaq}), which provides the iterative
equation:

\begin{equation}
 H^{\rm eff}_1 (\omega_n) = 
\sum_{m=0}^{\infty} \left[-P H_1 Q \left( \frac{-1}{\epsilon_0 -QHQ}
  \right)^{m+1} QH_1P \right] \left[ H^{\rm eff}_1 (\omega_{n-1})
\right]^m ~.
\label{eqa}
\end{equation}

The quantity inside the square brackets of Eq. (\ref{eqa}), that will
be dubbed from now on as $\hat{Q}_m(\epsilon_0)$, is proportional to
the $m$-th derivative of the \qbox~calculated in
$\epsilon=\epsilon_0$:

\begin{equation}
\hat{Q}_m(\epsilon_0) = -P H_1 Q \left( \frac{-1}{\epsilon_0 -QHQ}
  \right)^{m+1} QH_1P = \frac{1}{m!} \left[ \frac{d^m \hat{Q} 
(\epsilon)} {d \epsilon^m} \right]_{\epsilon=\epsilon_0} ~.
\label{eqb}
\end{equation}

We may then rewrite Eq. (\ref{eqa}), according to the above identity,
as:

\begin{equation}
 H^{\rm eff}_1 (\omega_n) =
\sum_{m=0}^{\infty} \frac{1}{m!} \left[ \frac{d^m \hat{Q} (\epsilon)} {d
    \epsilon^m} \right]_{\epsilon=\epsilon_0} \left[ H^{\rm eff}_1
  (\omega_{n-1}) \right]^m  = 
\sum_{m=0}^{\infty} \hat{Q}_m(\epsilon_0) \left[ H^{\rm eff}_1
  (\omega_{n-1}) \right]^m ~.
\label{eqc}
\end{equation}

The starting point of the KK iterative method is the assumption that $
H^{\rm eff}_1 (\omega_0)=\hat{Q} (\epsilon_0)$, which leads to
rewrite Eq. (\ref{eqc}) in the following form:

\begin{equation}
H^{\rm eff} = \sum_{i=0}^{\infty} F_i~, \label{kkeq}
\end{equation}

\noindent
where

\begin{eqnarray}
F_0 & = & \hat{Q}(\epsilon_0)  \nonumber \\
F_1 & = & \hat{Q}_1(\epsilon_0)\hat{Q}(\epsilon_0)  \nonumber \\
F_2 & = & \hat{Q}_2(\epsilon_0)\hat{Q}(\epsilon_0)\hat{Q}(\epsilon_0) + 
\hat{Q}_1(\epsilon_0)\hat{Q}_1(\epsilon_0)\hat{Q}(\epsilon_0)  \nonumber \\
~~ & ... & ~~ 
\label{kkeqexp}
\end{eqnarray}

The above expression represent the well-known folded-diagram expansion
of the effective Hamiltonian as introduced by Kuo and Krenciglowa,
since in Ref. \citep{Krenciglowa74} the authors demonstrated the
following operatorial identity: 

\begin{equation}
\hat{Q}_1\hat{Q}= - \hat{Q} \int \hat{Q}~,
\end{equation}

\noindent
where the integral sign corresponds to the so-called folding operation
as introduced by Brandow in Ref. \citep{Brandow67}.

\vspace{0.4truecm}
\subsubsection{The Lee-Suzuki Iterative Technique}\label{LS}
\vspace{0.1truecm}
The Lee-Suzuki (LS) technique is another iterative procedure which can
be carried out by rearranging Eq. (\ref{eqfinal}) in order to obtain
an explicit expression of the effective Hamiltonian $H^{\rm eff}_1$ in
terms of the operators $\omega$ and $\hat{Q}$ \citep{Suzuki80}:

\begin{equation} 
H^{\rm eff}_1 (\omega) = \left( 1 + P H_1 Q \frac{1}{\epsilon_0 - Q H Q}
  \omega \right)^{-1} \hat{Q} (\epsilon_0) ~.
\end{equation}

The iterative form of the above equation is the following:

\begin{equation} 
H^{\rm eff}_1 (\omega_n) = \left( 1 + P H_1 Q
\frac{1}{\epsilon_0 - Q H Q} \omega_{n-1} \right)^{-1} \hat{Q}
(\epsilon_0)~,
\end{equation}

\noindent
and we may also write an iterative expression of Eq. (\ref{omegaq}):

\begin{equation} 
\omega_n = Q \frac{1}{\epsilon_0 -QHQ} Q H_1 P - Q \frac{1}{\epsilon_0
    -QHQ} \omega_{n-1}H^{\rm eff}_1(\omega_n) ~.
\end{equation}

The standard procedure is to start the iterative procedure by choosing
$\omega_0=0$, so that we may write:

\begin{eqnarray}
H^{\rm eff}_1 (\omega_1) & = & \hat{Q} (\epsilon_0) \nonumber \\
\omega_1 & = &  Q \frac{1} {\epsilon_0 -Q H Q} Q H_1 P ~. \nonumber
\end{eqnarray}

It can be demonstrated, by performing some algebra, the following identity:

\begin{equation}
 \hat{Q}_1 (\epsilon_0) = - P H_1 Q \frac{1}{\epsilon_0 - QHQ} Q
\frac{1}{\epsilon_0 - QHQ} Q H_1 P  =  - P H_1 Q \frac{1}{\epsilon_0 - QHQ}
\omega_1 ~,
\end{equation}

\noindent
then, for the following iteration $n=2$ we have:

\begin{eqnarray}
H^{\rm eff}_1 (\omega_2) & = & \left( 1 + P H_1 \frac{1} {\epsilon_0 -
    QHQ} \omega_1 \right)^{-1} \hat{Q}(\epsilon_0) =  \nonumber \\
~ & = & \frac{1}{1 - \hat{Q}_1(\epsilon_0)} \hat{Q} (\epsilon_0) \nonumber \\
\omega_2 & = & Q \frac{1}{\epsilon_0 - QHQ} Q H_1 P -  Q
\frac{1}{\epsilon_0 -QHQ} \omega_{1}H^{\rm eff}_1(\omega_2)  ~.
\end{eqnarray}

Finally, the LS iterative expression of \heff~is the following:

\begin{equation}
H^{\rm eff}_1 (\omega_n) = \left[
1 - \hat{Q}_1 (\epsilon_0) \sum_{m=2}^{n-1} \hat{Q}_m (\epsilon_0)
\prod_{k=n-m+1}^{n-1} H^{\rm eff}_1 (\omega_k)
\right]^{-1} \hat{Q} (\epsilon_0)  ~~. \label{eqLS}
\end{equation}

It is important to point out that KK and LS iterative techniques,
which allow the solution of the decoupling equation \ref{deceq3}, in
principle do not provide the same \heff.
Suzuki and Lee have shown that the KK iterative approach provides an
effective Hamiltonian whose eigenstates have the largest overlap with
the model space ones, and that \heff~obtained employing the LS
technique has eigenvalues that are the lowest in energy among those
belonging to the set of the full Hamiltonian $H$ \citep{Suzuki80}.

Both procedures we have presented are are limited to
  employ an unperturbed Hamiltonian $H_0$ whose model-space
  eigenstates are degenerate in energy.
However, in Ref. \citep{Kuo95} the authors have introduced an
alternative approach to the standard KK and LS techniques, whose goal
is to extend these methods to the non-degenerate case by introducing
multi-energy $\hat{Q}$-boxes. 
Actually, this approach is quite involved for practical
applications, the only one existing in the literature being that in
Ref. \citep{Coraggio05c}.

In the following sections, we outline two methods
\citep{Takayanagi11a,Suzuki11} to derive effective SM Hamiltonians
which may be implemented straightforwardly to employ $H_0$s that are
non-degenerate within the model space.
\vspace{0.4truecm}

\subsubsection{The Kuo-Krenciglowa technique extended to
  non-degenerate model spaces}
\vspace{0.1truecm}
The extended Kuo-Krenciglowa (EKK) method is an extension of the KK
iterative technique to derive a \heff~ within non-degenerate model
spaces \citep{Takayanagi11a,Takayanagi11b}.

We will now summarize the EKK method as follows.

\noindent
First, a shifted Hamiltonian $\tilde{H}$ is introduced in terms of an
energy parameter $E$
\begin{equation}\label{Htilde}
\tilde{H}=H-E~.
\end{equation}
\noindent
Then, we rewrite Eq. (\ref{deceq3}) in terms of $\tilde{H}$:

\begin{equation}
 (E - QHQ) \omega = QH_1P - \omega P\tilde{H}P - \omega PH_1Q \omega =
 QH_1P - \omega \tilde{H}_{\rm eff} ~.\label{EKKdec}
\end{equation}
\noindent
Eq. (\ref{EKKdec}) may be solved by way of an iterative procedure,
analogously to the KK technique, in terms of the \qbox~and its
derivatives, as defined in Eqs. (\ref{qbox}) and (\ref{eqb}),
respectively.

The effective Hamiltonian $\tilde{H}_{\rm eff}$ at the $n$-step of the
iterative procedure may be then expressed as follows
\citep{Takayanagi11a}:

\begin{equation}
  \tilde{H}_{\rm eff}^{(n)}= \tilde{H}_{\rm BH}(0) + \sum_{k=1}^{\infty}
  \hat{Q}_k (0) \left[ \tilde{H}_{\rm eff}^{(n-1)} \right]^k~, \label{EKKHeffn}
\end{equation}

\noindent
where $\tilde{H}_{\rm BH}$ is the solution of the Bloch-Horowitz
equation \citep{Bloch58}:

\begin{equation}
  \tilde{H}_{\rm BH} (E)=  P\tilde{H}P+PH_1Q \frac{1}{E-QHQ} Q H_1 P
  ~.\label{eqblochhor}
\end{equation}

We observe that the EKK method does not require $H_0$ to be degenerate
within the model space, and has been therefore applied to derive \heff~in
a multi-shell valence space \citep{Tsunoda14a,Tsunoda17} and in Gamow SM
calculations with realistic $NN$ potentials \citep{Sun17,Ma20}.

It is worth pointing out that, since $\tilde{H}_{\rm
  eff}=\lim\limits_{n \to \infty} \tilde{H}_{\rm  eff}^{(n)}$, we can
write

\begin{equation}
  \tilde{H}_{\rm eff}= \tilde{H}_{\rm BH}(0) + \sum_{k=1}^{\infty}
  \hat{Q}_k (0) \left[ \tilde{H}_{\rm eff} \right]^k~, \label{EKKtHeff}
\end{equation}
\noindent
Eq. (\ref{EKKtHeff}) may be interpreted as a Taylor series expansion of
$\tilde{H}_{\rm eff}$ around $\tilde{H}_{\rm BH}$, and the parameter
$E$ corresponds to a shift of the origin of the expansion, and a
resummation of the series \citep{Tsunoda14a}.
As a matter of fact, because of Eq. (\ref{Htilde}) we may express \heff~as

\begin{equation}
H_{\rm eff} = \tilde{H}_{\rm eff}+E= {H}_{\rm BH}(0) + \sum_{k=1}^{\infty}
  \hat{Q}_k (0) \left[ \tilde{H}_{\rm eff} \right]^k~, \label{EKKHeff}
\end{equation}

Now, both sides of the above equation will be independent of $E$,
providing that the summation is carried out at infinity, and the
parameter $E$ may be tuned to accelerate the convergence of the
series, when in practical applications a numerical partial summation
needs to be employed and a perturbative expansion
of the \qbox~is carried out \citep{Tsunoda14a}.
\vspace{0.4truecm}

\subsubsection{The $\hat{Z}(\epsilon)$ vertex function}\label{ZB}
\vspace{0.1truecm}
Suzuki and coworkers in Ref. \citep{Suzuki11} have proposed an
approach to the derivation of \heff~ that aims to avoid the
divergencies of the \qbox~vertex function, if a non-degenerate model
space in considered.
In fact, the definition of the \qbox~in Eq. (\ref{qbox}) evidences
that if $\epsilon$ approaches one of the eigenvalues of $QHQ$, then
instabilities may arise if one employs a numerical derivation, since
these eigenvalues are poles for $\hat{Q}(\epsilon)$.

We now sketch out the procedure of Ref. \citep{Suzuki11} and, for the
sake of simplicity, consider the case of a degenerate unperturbed
model space (i.e., $ P H_0 P =\epsilon_0 P $).

A new vertex function $\hat{Z}(\epsilon)$ is introduced and defined in
terms of $\hat{Q}(\epsilon)$ and its first derivative as follows:

\begin{equation}
\label{eq:z-box}
 \hat Z(\epsilon) \equiv \frac{1}{{1 - \hat Q_1 (\epsilon)}}\left[
   {\hat Q(\epsilon) - \hat Q_1 (\epsilon) (\epsilon - \epsilon_0 ) P
   } \right]~.
 \end{equation}

It can be demonstrated that $\hat Z(\epsilon)$ satisfies the following
equation \citep{Suzuki11}:

\begin{equation}
\left[ \epsilon_0  + \hat Z(E_\alpha) \right] P | \Psi_{\alpha}
\rangle  = E_{\alpha} P | \Psi_{\alpha} \rangle ~~~~~~~~~~(\alpha=1,..,d)~.
\end{equation}

\noindent
Consequently, $H^{\rm eff}_1$ may be obtained by calculating the
$\hat{Z}$-box for those values of the energy, determined
self-consistently, that correspond to the ``true'' eigenvalues ${E_\alpha}$.

To calculate ${E_\alpha}$, we solve the following eigenvalue problem 
\begin{equation} 
 \left[ {\epsilon_0  + \hat Z(\epsilon)} \right] | {\phi_k
 }\rangle=F_{k}(\epsilon) |{\phi_k}\rangle~~,~~~~~~~~~~ (k = 1,2,\cdots,d)~~, 
\label{eq-Z-eigenvalue}
 \end{equation}

\noindent
where $F_{k}(\epsilon)$ are $d$ eigenvalues that depend on $\epsilon$.
Then, the true eigenvalues ${E_\alpha}$ can be obtained through the
solution of the $d$ equations

\begin{equation}
\label{eq:EFK}
\epsilon=F_k(\epsilon), \,\,\,\,(k=1,2,\cdots ,d)~~.
\end{equation}

First, it is worth pointing out some fundamental properties of $\hat
Z(\epsilon)$ and of the associated functions $F_k(\epsilon)$ and then
we proceed to discuss the solution of the equations
(\ref{eq-Z-eigenvalue},\ref{eq:EFK}). 

The behavior of $\hat Z(\epsilon)$ in proximity of the poles of $\hat
Q(\epsilon)$ is dominated by $\hat Q_1(\epsilon)$, and we may write
$\hat{Z}(\epsilon) \approx (\epsilon - \epsilon_0)P$.
This means that $\hat{Z}(\epsilon)$ has no poles and therefore
$F_k(\epsilon)$ are continuous and differentiable functions for any
value of $\epsilon$.

Eqs. (\ref{eq:EFK}) may have solutions that do not correspond to the
true eigenvalues ${E_\alpha}$, namely spurious solutions. 
In Ref. \citep{Suzuki11} it has been shown that, since the energy
derivative of $F_k(\epsilon)$ approaches to zero at $\epsilon = E_\alpha$,
the study of this derivative provides a criterion to locate and reject
spurious solutions.
The solution of Eqs. (\ref{eq-Z-eigenvalue}) and (\ref{eq:EFK}), that
are necessary to derive the effective interaction, may be obtained
employing both iterative and non-iterative methods.

We describe here a graphical non-iterative method to solve
Eqs. (\ref{eq:EFK}).

\noindent
As mentioned before, the $F_k(\epsilon)$'s are continuous functions of
the energy, therefore the solutions of
Eqs. (\ref{eq:EFK}) may be determined as the intersections of the
graphs $y = \epsilon$ and $y = F_k(\epsilon)$ employing one of the
well-known algorithms to solve nonlinear equations.

More precisely, if we define the functions $f_k(\epsilon)$ as
$f_k(\epsilon) = F_k(\epsilon) - \epsilon$, the solutions of
Eqs. (\ref{eq:EFK}) can obtained by finding the roots of the equations
$f_k(\epsilon) = 0$.
From inspection of the graphs $y = \epsilon$ and $y = F_k(\epsilon)$,
we can locate for each intersection a small surrounding interval
$[\epsilon_a , \epsilon_b] $ where $f_k(\epsilon_a) f_k(\epsilon_b) <
0 $.
The assumption that $f_k(\epsilon)$ is a monotone function within this
interval implies the existence of a unique root, which
  can be accurately determined by means of the secant method algorithm (see for
instance Ref. \citep{recipes_for}).

After we have determined the true eigenvalues ${E_\alpha}$, the
effective Hamiltonian $H^{\rm eff}_1$ is constructed as
\begin{equation}
\label{eq:REKK}
 H^{\rm eff}_1 = 
 \sum\limits_{\alpha = 1}^d {\hat Z} (E_\alpha )|{\phi_\alpha}\rangle 
 \langle {\tilde\phi_\alpha}| ~~,
\end{equation}

\noindent
where $|{\phi_\alpha}\rangle $ is the eigenvector obtained from
Eq. (\ref{eq-Z-eigenvalue}) while  $\langle {\tilde\phi_\alpha}|$ is
the correspondent biorthogonal state ($\langle{\tilde\phi_{\alpha} |\phi_{\alpha'}
}\rangle  = \delta_{\alpha \alpha'}$).

As we have mentioned at the beginning of this section,  we have
considered the case of a degenerate unperturbed model space (i.e., $ P
H_0 P =\epsilon_0 P $), but the above formalism can be easily
generalized to the non-degenerate case replacing $\epsilon_0 P $ with
$ P H_0 P $ in Eqs. (\ref{eq:z-box}-\ref{eq-Z-eigenvalue}).
\vspace{0.4truecm}

\subsection{The diagrammatic expansion of the $\hat{Q}$-box vertex function}
\label{qboxsec}
\vspace{0.1truecm}
The methods to derive \heff, which have been presented in the previous
sections, need the calculation of the \qbox~function vertex
function:

\begin{equation}
\hat{Q} (\epsilon) = P H_1 P + P H_1 Q \frac{1}{\epsilon - Q H Q} Q
H_1 P ~. \nonumber
\end{equation}

For our purpose, the term $1/(\epsilon - Q H Q)$ should be expanded as
a power series

\begin{equation}
\frac{1}{\epsilon - Q H Q} = \sum_{n=0}^{\infty} \frac{1}{\epsilon -Q
  H_0 Q} \left( \frac{Q H_1 Q}{\epsilon -Q H_0 Q} \right)^{n} ~,
\end{equation}

\noindent
leading to a perturbative expansion of the \qbox.
It is useful to employ a diagrammatic representation of this
perturbative expansion, which is a collection of Goldstone diagrams
that have at least one $H_1$-vertex, are irreducible - namely at least
one line between two successive vertices does not belong to the model
space - and are linked to at least one external valence line (valence
linked) \citep{Kuo71}.

The standard procedure for most perturbative derivations of \heff~is
to deal with systems with one and two valence nucleons, but later we
will also show our way to include contributions from three-body
diagrams that come into play when more than two valence nucleons are
considered.
$H_{\rm eff}^{1b}$ of single valence-nucleon nuclei provides the theoretical
effective SP energies, while TBMEs of the residual interaction $V^{\rm
  eff}$ are obtained from the $H_{\rm eff}^{2b}$ for systems with two
valence nucleons.
This can be achieved by a subtraction procedure \citep{Shurpin83},
namely removing from $H_{\rm eff}^{2b}$ the diagonal component of the
effective SP energies, derived from the $H_{\rm eff}^{1b}$ of the one
valence-nucleon systems.

A useful tool, for those who want acquire a sufficient knowledge about
the calculation of \qbox~diagrams in an angular momentum coupled
representation, is the paper by Kuo and coworkers in Ref. \citep{Kuo81}.

It is worth pointing out that in current literature the effective SM
Hamiltonians are derived accounting for \qbox~diagrams at most up to
the third order in perturbation theory, since it is computationally
highly demanding to perform calculations including
complete higher-order sets of diagrams.
A complete list can be found in Ref. \citep{Coraggio12a}, Appendix B,
and it is a collection of 43 one-body and 135 two-body diagrams.
it should be pointed out that lists of diagrams may be easily obtained
using algorithms which generate order-by-order Hugenholtz diagrams for
perturbation theory applications (see, for example,
Ref. \citep{Paldus73}).

\begin{figure}[h!]
\begin{center}
\includegraphics[width=10cm]{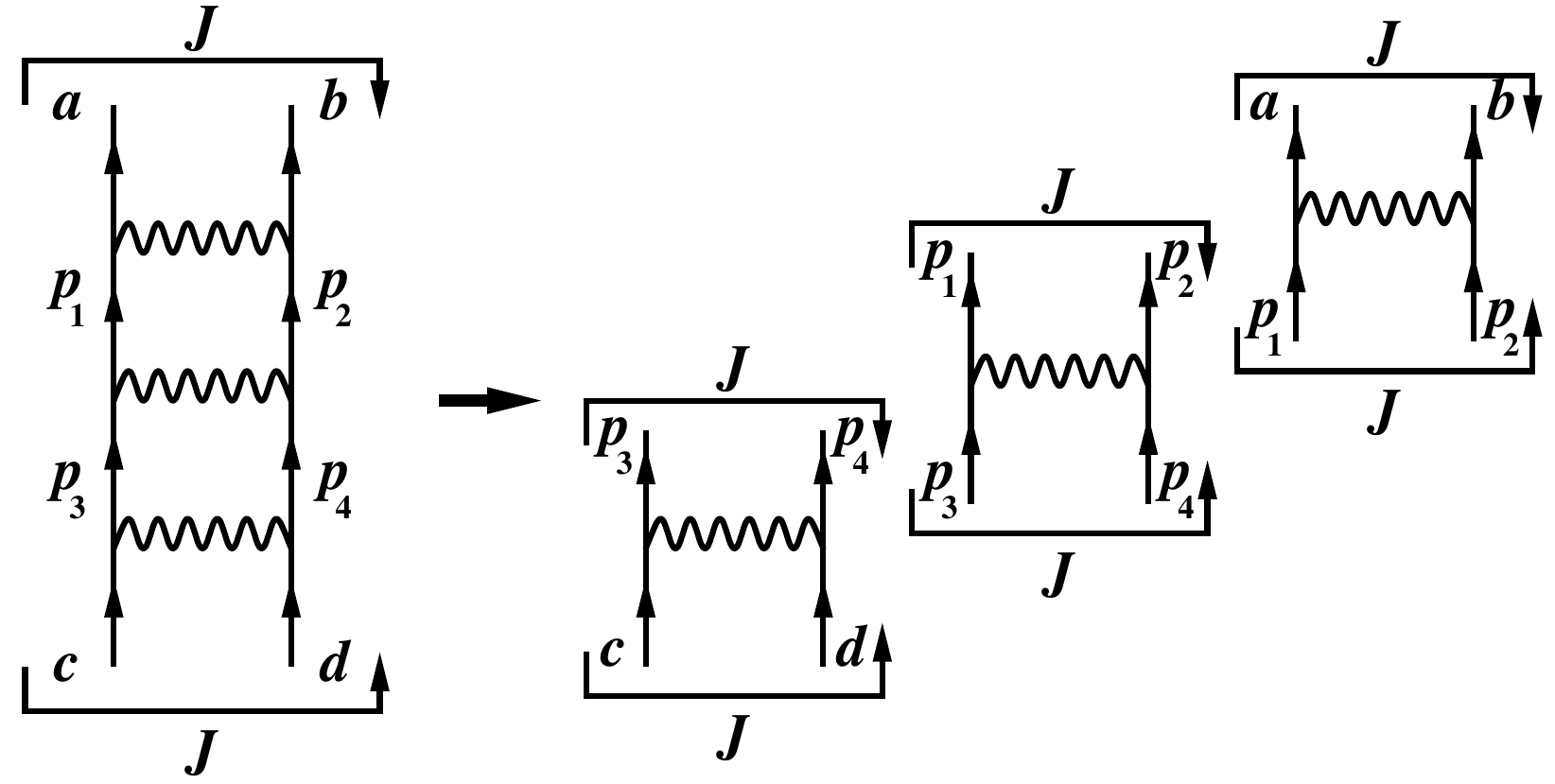}
\end{center}
\caption{Two-body ladder diagram at third order in perturbation
  theory. Arrow lines represent incoming/outcoming and intermediate
  particle states. Wavy lines indicate interaction vertices.}\label{ladderd}
\end{figure}

Since the aim of present work is to support practitioners with useful
tips to derive effective SM Hamiltonians within the perturbative
approach, we are going to show some selected examples of
\qbox~diagrams and their analytical expression.
Our first example is the third-order ladder diagram $V_{\rm ladder}$
shown in Fig. \ref{ladderd}, and to explicit its expression we will use the
proton-neutron angular-momentum coupled representation for the TBMEs of
the input $V_{NN}$:

\begin{equation}\label{tbmepn}
\langle 1,2; J| V_{NN} | 3,4; J\rangle \equiv
\langle n_1l_1j_1t_{z_1},n_2l_2 j_2t_{z_2}; J| V_{NN} |n_3l_3 j_3t_{z_3},
n_4l_4j_4t_{z_4}; J\rangle~.
\end{equation}

The TBMEs elements of the input potential $V_{NN}$ are antisymmetrized
but not normalized to ease the calculation of the \qbox~diagrams,
$n_m,l_m,j_m,t_{z_m}$ indicate the orbital and isospin quantum numbers
of the SP state $m$.

The analytical expression of $V_{\rm ladder}$ is:

\begin{equation}\label{laddere}
\langle a,b; J| V_{\rm ladder} | c,d; J\rangle =+\frac{1}{4}
\sum_{p_1p_2p_3p_4}\frac{
\langle a,b; J| V_{NN} | p_1,p_2; J\rangle
\langle p_1,p_2; J| V_{NN} | p_3,p_4; J\rangle
\langle p_3,p_4; J| V_{NN} | c,d; J\rangle
}
{
[\epsilon_{0}-(\epsilon_{p_1}+\epsilon_{p_2})]
[\epsilon_{0}-(\epsilon_{p_3}+\epsilon_{p_4})]
}~,
\end{equation}

where $\epsilon_m$ denotes the unperturbed single-particle energy of
the orbital $j_m$, $\epsilon_{0}$ is the so-called starting energy,
namely the unperturbed energy of the incoming particles
$\epsilon_0=\epsilon_{c}+\epsilon_{d}$.

We point out that the factor $+1/4$ is related to the rules that
characterize the calculation of overall factors in \qbox~Goldstone
diagrams; for any diagram we have a phase factor
\[
  (-1)^{(n_h+n_l+n_c+n_{\rm exh})}
\]
whose value is determined by the total number of hole lines ($n_h$),
the total number of closed loops ($n_l$), the total number of
crossings of different external lines as they trace through the
diagrams ($n_c$), and the total number of external hole lines which
continuously trace through the diagrams ($n_{\rm exh}$)
\citep{Kuo81}.
There is also a factor $(1/2)^{n_{\rm ep}}$, that accounts of the number
of pairs of lines which start together from one interaction vertex and
end together to another one ($n_{\rm ep}$).

The diagram in Fig. \ref{ladderd} exhibits $n_h=n_l=n_c=n_{\rm
  exh}=0$, consequently the phase is positive.
The number of pairs of particles starting and ending together in the
same vertices is $n_{\rm ep}=2$, and consequently the overall factor
is $+1/4$.

\begin{figure}[h!]
\begin{center}
\includegraphics[width=17cm]{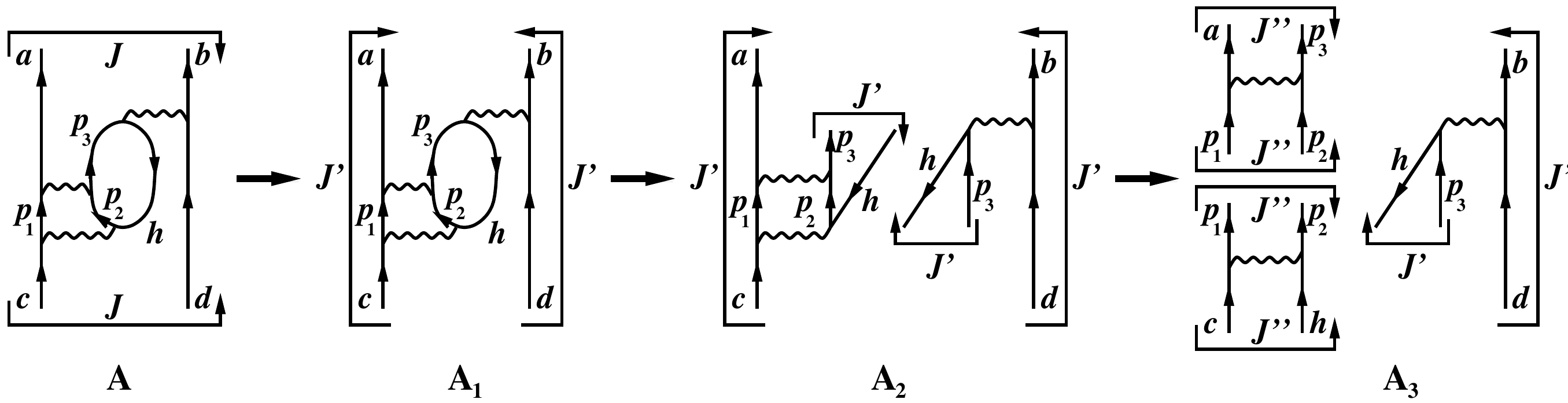}
\end{center}
\caption{Two-body $3p\mbox{-}1h$ diagram at third order in perturbation
  theory. Arrow lines represent incoming/outcoming and intermediate
  particle/hole states. Wavy lines indicate interaction vertices.}\label{g3p1hd}
\end{figure}

The factorization of Goldstone diagrams such as the ladder one in
Fig. \ref{ladderd} is quite simple in terms of its interaction
vertices.
There is a large class of diagrams, as for example the
three-particle-one-hole diagram ($3p\mbox{-}1h$) in Fig. \ref{g3p1hd},
which require some considerations in order to provide a straightforward
factorization.

The factorization may be easily performed by taking into account that the
interaction operator $V_{NN}$ transforms as a scalar under rotation,
and so introducing the following cross-coupling transformation of the
TBMEs:

\begin{equation}\label{cc}
\langle a,b; J| V_{NN} | c,d; J\rangle_{\rm{\bf CC}} = \frac{1}{\hat{J}} \sum_{J'}
\hat{J'} X \left( \begin{array}{ccc}
j_c~ j_a~  J  \\ 
j_d~ j_b~  J  \\ J'~  J'~  0  \end{array} \right)
\langle a,b; J'| V_{NN} | c,d; J'\rangle~,
\end{equation}

\noindent
where $\hat{x}=(2x+1)^{1/2}$.
$X$ is the so-called standard normalized 9-$j$ symbol, expressed as
follows in terms of the Wigner $9\mbox{-}j$ symbol \citep{Edmonds57}:

\[
X \left( \begin{array}{ccc}
r ~ s ~  t  \\ 
u~ v ~  w  \\ x~  y~  z  \end{array} \right) = \hat{t} \hat{w} \hat{x} \hat{y}
\left\{ \begin{array}{ccc}
r ~ s ~  t  \\ 
u~ v ~  w  \\ x~  y~  z  \end{array} \right\}~.
\]

The orthonormalization properties of the $X$ symbol allow then to write the
direct-coupled TBMEs in terms of the cross-coupled TBMEs:

\begin{equation}\label{cc_inv}
\langle a,b; J| V_{NN} | c,d; J\rangle = \frac{1}{\hat{J}} \sum_{J'}
\hat{J'} X \left( \begin{array}{ccc}
j_c~ j_d~  J  \\ 
j_a~ j_b~  J  \\ J'~  J'~  0  \end{array} \right)
\langle a,b; J'| V_{NN} | c,d; J'\rangle_{\rm{\bf CC}}
\end{equation}

Eqs. (\ref{cc},\ref{cc_inv}) help to perform the factorization of
diagram in Fig. \ref{g3p1hd}; first a rotation according
Eq. (\ref{cc_inv}) transforms the direct coupling to the total angular
momentum $J$ into the cross-coupled one $J'$(diagram A going into
diagram ${\rm A}_1$ in Fig. \ref{g3p1hd}).
This allows to cut the inner loop and factorize the diagram into two
terms, a ladder component ($\alpha$) and a cross-coupled matrix element
($\beta$) (diagram ${\rm A}_2$ in Fig. \ref{g3p1hd}):

\begin{eqnarray}
(\alpha) & = & \langle a,p_3; J'| \mathrm{A} | c,h; J'\rangle_{\rm{\bf CC}}
                   \nonumber \\ 
(\beta) & = & \langle h,b; J'| V_{NN} | p_3,d; J'\rangle_{\rm{\bf CC}} \nonumber
\end{eqnarray}

Then, we transform the ladder diagram (A) back to a direct coupling to
$J''$ by way of Eq. (\ref{cc}), and factorize it into the TBMEs (I)
and (II) (diagram ${\rm A}_3$ in Fig. \ref{g3p1hd}):

\begin{eqnarray}
\mathrm{(I)}~ & = & \langle a,p_3; J''| V_{NN} | p_1,p_2; J''\rangle
                   \nonumber \\
\mathrm{(II)} & = & ~ \langle p_1,p_2; J''| V_{NN} | c,h; J''\rangle \nonumber
\end{eqnarray}

The analytical expression of the diagram in Fig. \ref{g3p1hd} is the following:

\begin{eqnarray}\label{g3p1he}
\langle a,b; J| V_{3p1h} | c,d; J\rangle&=&-\frac{1}{2}\frac{1}{\hat{J}}
  \sum_{hp_1p_2p_3}\sum_{J'J''} \hat{J''} X \left( \begin{array}{ccc}
j_c~ j_d~  J  \\ 
j_a~ j_b~  J  \\ J'~  J'~  0  \end{array}\right)
  X \left( \begin{array}{ccc}
j_c~ j_a~  J'  \\ 
j_h~ j_p~  J'  \\ J''~  J''~  0  \end{array} \right) \\
~ & \times &\frac{
\langle h,b; J'| V_{NN} | p_3,d; J'\rangle_{\rm{\bf CC}}
\langle a,p_3; J''| V_{NN} | p_1,p_2; J''\rangle
\langle p_1,p_2; J''| V_{NN} | c,h; J''\rangle
}
{
[\epsilon_{0}-(\epsilon_{p_1}+\epsilon_{p_2})]
[\epsilon_{0}-(\epsilon_{p_3}+\epsilon_{p_4})]
}~,\nonumber
\end{eqnarray}

The factor $(-1/2)$ accounts the fact that $n_{\rm ep}=1$,
$n_h=n_l=1$, and that an extra-phase factor $(-1)^{n_{\rm ph}}$ is
needed for the total number of cut of particle-hole pairs ($n_{\rm
  ph}$) \citep{Kuo81}, since in order to factorize the diagram we have
cut the inner loop.

It is important pointing out that there are other three diagrams with
the same topology as the one in Fig. \ref{g3p1hd}, which corresponds
to the exchange of the external incoming and outcoming particles.

Let us now turn our attention to one-body diagrams.

\begin{figure}[h!]
\begin{center}
\includegraphics[width=7cm]{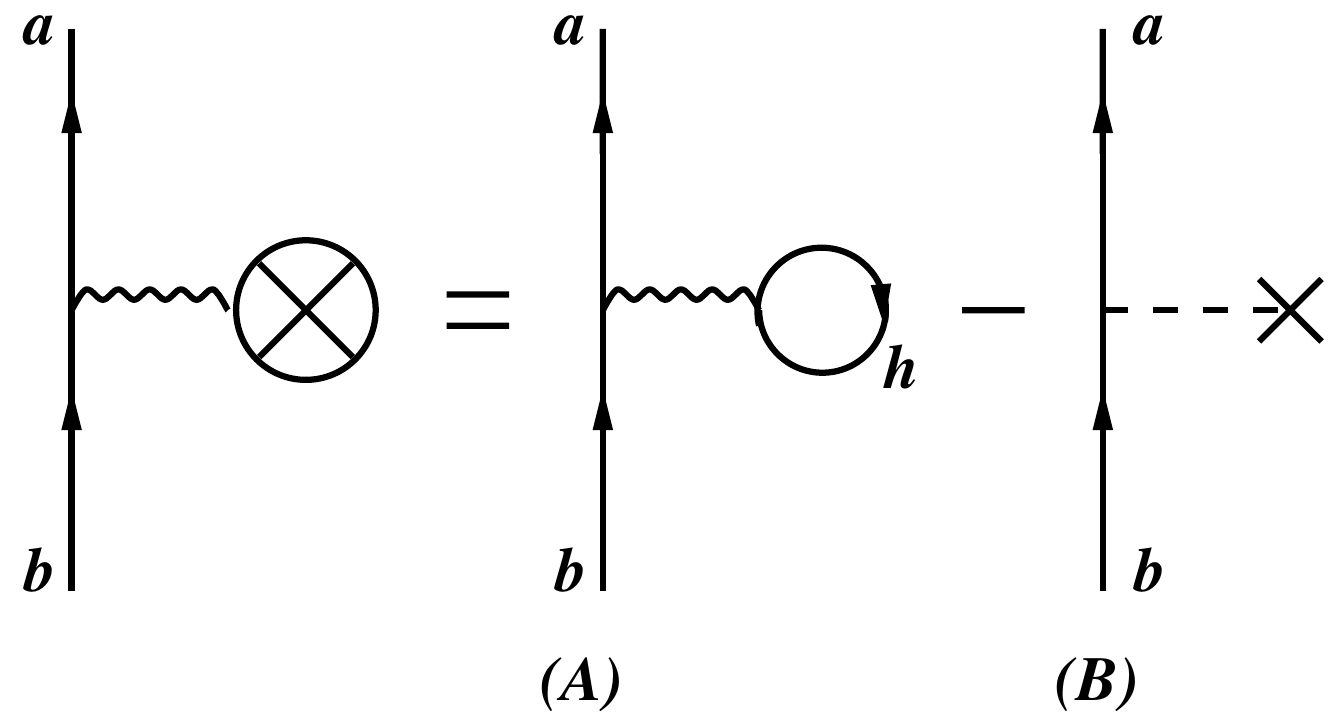}
\end{center}
\caption{($V\mbox{-}U$)-insertion diagram. Graph ($A$) is the
  self-energy diagram. Graph ($B$) represents the matrix element of
  the harmonic oscillator potential $U=\frac{1}{2}m \omega^2 r^2$.}
\label{VUd}
\end{figure}

First of all, we consider the contribution of diagrams such as the one
in Fig. \ref{VUd}.

The diagram in Fig. \ref{VUd} is the so-called $(V\mbox{-}U)$-insertion
diagram, and is composed of the self-energy diagram ($V$-insertion
diagram) minus the auxiliary potential $U$-insertion.
The $U$-insertion diagrams are due to the presence of the
  $\mbox{-}U$ term in $H_1$.
The analytical expression of this diagrams is the following:

\begin{eqnarray}\label{VUe}
\langle a || (V\mbox{-}U) || b \rangle & = &
\frac{\delta_{j_aj_b}}{2j_a+1} \sum_{Jh} (2J +1) \langle j_a, h;J | V |
j_b,h;J \rangle - \langle a || U || b \rangle \\
~ & = &
\frac{\delta_{j_aj_b}}{2j_a+1} \sum_{Jh} (2J +1) \langle j_a, h;J | V |
j_b,h;J \rangle - \langle a || \frac{1}{2}m \omega^2 r^2|| b \rangle \nonumber
\end{eqnarray}

The calculation of the self-energy diagram $A$ has been performed by
coupling the external lines to a scalar, which leads the SP total angular
momentum and parity $j_a,j_b$ being identical.
Then, we cut the inner hole line and, since SP states $a,b$ are coupled to
$J=0^+$, we apply the transformation in Eq. (\ref{cc}) for $J=0^+$.

Since the standard choice for the auxiliary potential is the
harmonic-oscillator (HO) one, it appears also the reduced matrix
element of $U=\frac{1}{2}m \omega^2 r^2$ between SP states $a$ and
$b$ (graph $B$).

It is worth pointing out that the diagonal contributions of
($V\mbox{-}U$)-insertion diagrams, for SP states belonging to the model
space, correspond to the first order contribution of the perturbative
expansion of the effective SM Hamiltonian of single valence-nucleon
systems $H_{\rm eff}^{1b}$.

Moreover, $(V\mbox{-}U)$-insertion diagrams turn out to be identically
zero when employing a self-consistent Hartree-Fock (HF) auxiliary
potential \citep{Coraggio05c}, and in Ref. \citep{Coraggio12a} it has
been discussed the important role played by these terms, comparing
different effective Hamiltonians derived starting from $\hat{Q}$-boxes
with and without contributions from $(V\mbox{-}U)$-insertion diagrams. 

Now, we will show an example of one-body diagram and comment briefly
its analytical calculation.

\begin{figure}[h!]
\begin{center}
\includegraphics[width=14cm]{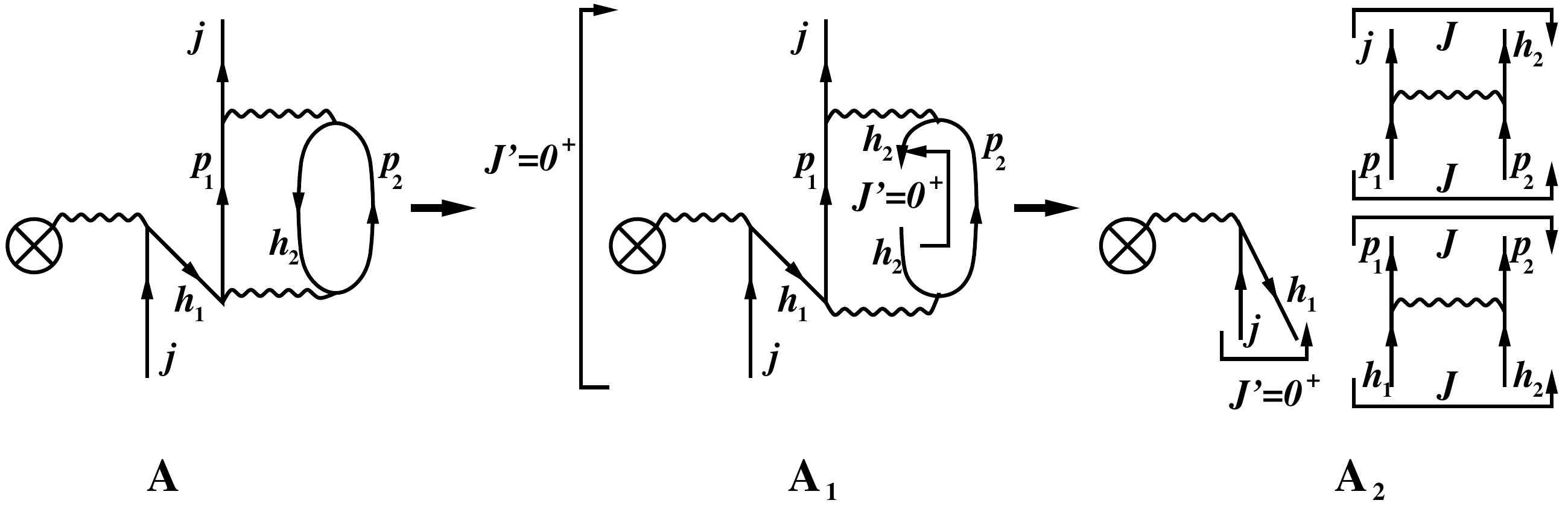}
\end{center}
\caption{An example of one-body diagram (see text for details).}
\label{d1hf20d}
\end{figure}

We consider the diagram as reported in Fig. \ref{d1hf20d}, while the
complete list of third-order one-body diagrams can be found in
Ref. \citep{Coraggio12a} Fig. B.19.

We dub this diagram $V_{2p1h}$, since between the upper interaction
vertices they appear two particles and 1 hole as intermediate states.
This belongs to the group of non-symmetric diagrams, which occur
always in pairs giving equal contributions.
Its analytical expression is:

\begin{equation}\label{d1hf20e}
\langle j || V_{2p1h} || j \rangle = -\frac{1}{2}\frac{1}{2j+1}
\sum_{\substack{Jp_1p_2{}\\h_1h_2}} (2J+1) \frac{
\langle j,h_2; J| V_{NN} | p_1,p_2; J\rangle
\langle p_1,p_2; J| V_{NN} | h_1,h_2; J\rangle
\langle h_1|| V\mbox{-}U || j \rangle
}
{
[\epsilon_{0}-(\epsilon_{p_1}+\epsilon_{p_2}-\epsilon_{h_2})]
[\epsilon_{0}-(\epsilon_{j}+\epsilon_{p_1}+\epsilon_{p_2}-\epsilon_{h_1}-\epsilon_{h_2})]
}~,
\end{equation}

\noindent
$\epsilon_0=\epsilon_j$ being the unperturbed SP energy of the
incoming particle $j$.

In order to factorize the diagram, we have first cross-coupled the
incoming and outcoming model-space states $j$ to $J'=0^+$(diagram
${\rm A}_1$ in Fig. \ref{d1hf20d}).
Then we cut the hole-line $h_2$ and, by way of Eq. (\ref{cc}), we obtain
a sum of two-body diagrams which are direct-coupled to the total
angular momentum $J$ \citep{Kuo81} (diagram ${\rm A}_2$ in
Fig. \ref{d1hf20d}).
These operations are responsible of the factors $1/(2j+1)$ and
$(2J+1)$, the overall factor $1/2$ is due to the pair of particle
lines $p_1,p_2$ starting and ending at same vertices, and the minus
sign comes from the 2 hole lines and 1 loop appearing in the diagram.
The factorization accounts also of the $(V\mbox{-}U)$ insertion $\langle
h_1|| V\mbox{-}U | j \rangle$.

As mentioned before, this diagrammatics is valid to derive \heff for
one- and two-valence nucleon systems, and things are different and
more complicated if one would like to derive \heff for system with
three or more valence nucleons.

Actually, none of available SM codes can perform the diagonalization
of SM Hamiltonians with three-body components, apart from BIGSTICK SM
code \citep{BIGSTICK} but only for light nuclei.

\begin{figure}[h!]
\begin{center}
\includegraphics[width=6cm]{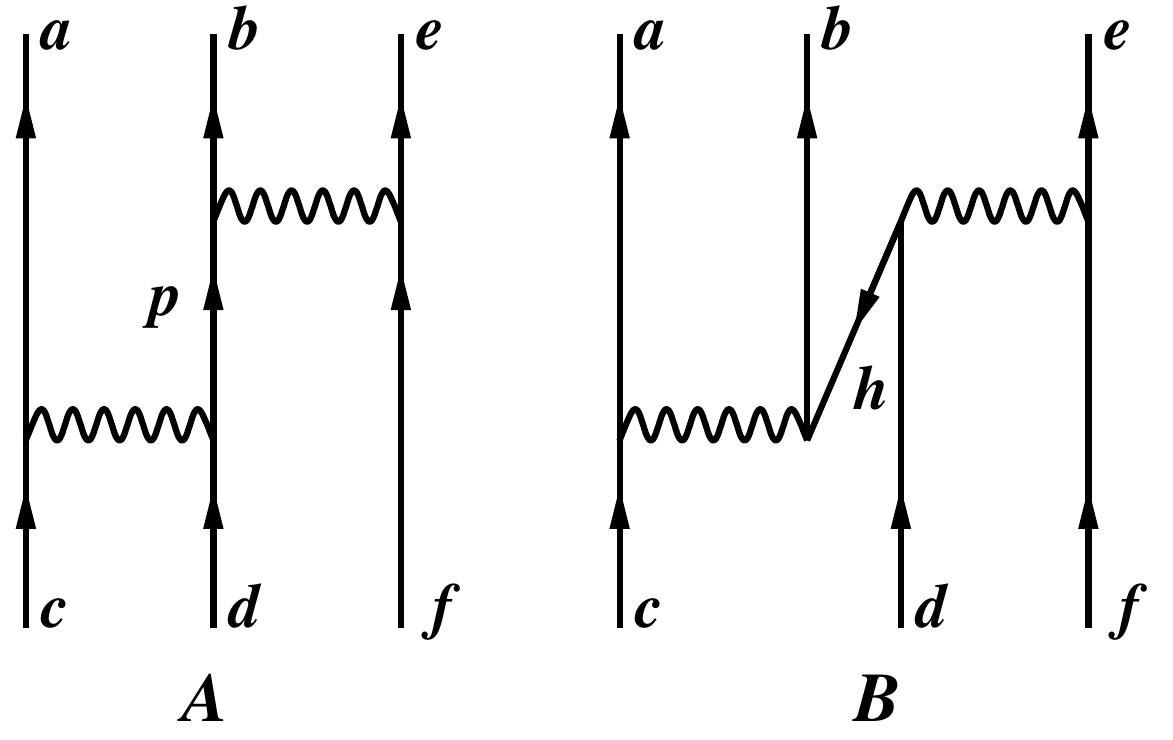}
\caption{Second-order three-body diagrams. The sum over the
  intermediate lines runs over particle and hole states outside the
  model space, shown by A and B, respectively. For the sake of
  simplicity, for each topology we report only one of the diagrams
  which correspond to the permutations of the external lines.}
\label{diagram3corr}
\end{center}
\end{figure}

In order to include the contribution to \heff~from \qbox~diagrams
with at least three incoming and outcoming valence particles, we
resort to the so-called normal-ordering decomposition of the
three-body component of a many-body Hamiltonian
\citep{HjorthJensen17}.
To this end, we include in the calculation of the $\hat{Q}$-box also
second-order three-body diagrams, which, for those nuclei with more
than 2 valence nucleons, account for the interaction via the two-body
force of the valence nucleons with core excitations as well as with
virtual intermediate nucleons scattered above the model space (see
Fig. \ref{diagram3corr}).

For each topology reported in Fig. \ref{diagram3corr}, there are nine
diagrams, corresponding to the possible permutations of the external
lines.
The analytical expressions of the second-order three-body
contributions is reported in Ref. \citep{Polls83}, and we derive from
those expressions a density-dependent two-body term.

\begin{figure}[h!]
\begin{center}
\includegraphics[width=5.5cm]{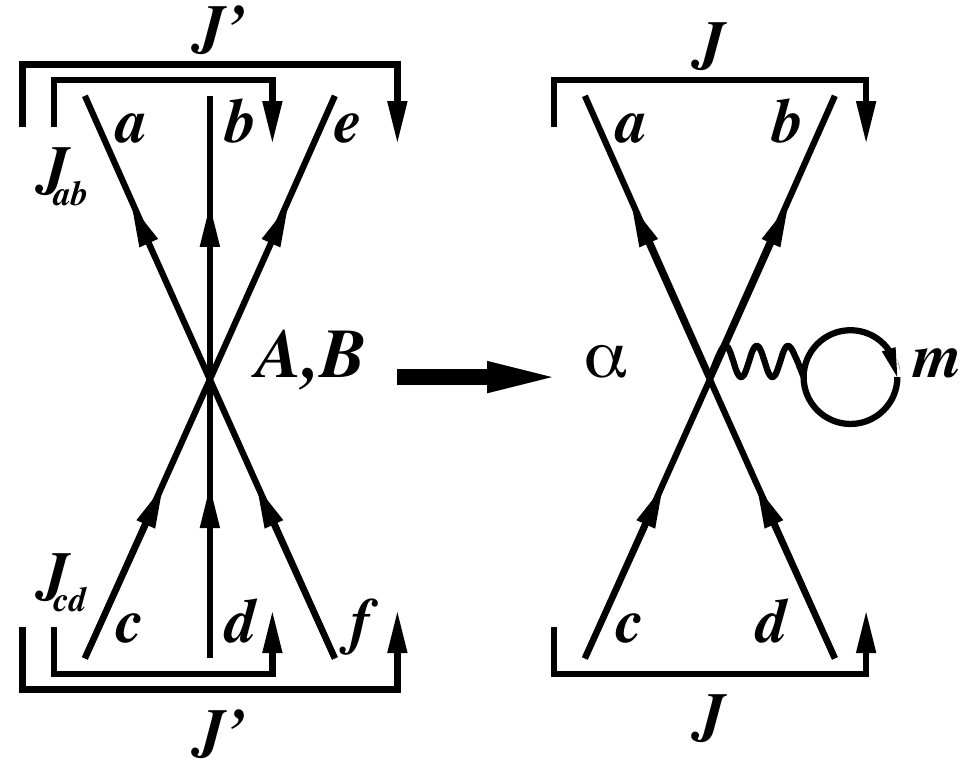}
\caption{Density-dependent two-body contribution that is
  obtained from a three-body one. $\alpha$ is obtained by
  summing over one incoming and outgoing particle of the three-body
  graphs $A$ reported in Fig. \ref{diagram3corr}.}
\label{3bf}
\end{center}
\end{figure}

To this end, we calculate, for each $(A,B)$ topology, nine one-loop
diagrams, namely the graph $(\alpha)$ in Fig. \ref{3bf}.
Their explicit form, in terms of the three-body graphs $(A,B)$, is:

\begin{equation}
\label{correq}
\langle (j_a j_b)_J| V^{\alpha} | (j_c j_d)_J \rangle =
\sum_{m,J'} ~ \rho_m \frac{\hat{J'}^2}{\hat{J}^2} \langle \left[
  (j_{a} j_{b})_{J},j_m \right]_{J'} | V^{A,B}  | \left[
  (j_{c} j_{d})_{J},j_m \right]_{J'} \rangle~~,
\end{equation}

\noindent
where the summation over $m$-index runs in the model space.
$\rho_m$ is the unperturbed occupation density of the orbital $m$
according to the number of valence nucleons.

Finally, the perturbative expansion of the $\hat{Q}$-box
  contains one- and two-body diagrams up to third order in $V_{NN}$, and a
density-dependent two-body contribution accounting for three-body
second-order diagrams \citep{Ellis77,Polls83}.

It should be pointed out that the latter term depends on the number of
valence protons and neutrons, thus leading to the
derivation of specific effective shell-model Hamiltonians, that
differ only for the two-body matrix elements.
\vspace{0.4truecm}

\subsection{Effective shell-model decay operators}\label{effopsec}
\vspace{0.1truecm}
In the shell-model approach, we are interested not only in calculating
energies, but also the matrix elements of operators $\Theta$ representing
physical observables (e.g. e.m. transition rates, multipole moments,
...).

Since the wave-functions $| \psi_{\alpha} \rangle$ obtained
diagonalizing \heff~are not the true ones $| \Psi_{\alpha} \rangle$,
but their projections onto the chosen model space
($|\psi_{\alpha} \rangle = P | \Psi_{\alpha} \rangle $), it is obvious
that one has to renormalize $\Theta$ to take into account
the neglected degrees of freedom corresponding to the $Q$-space.
In other words, one has to consider the short range correlation
“wounds” inflicted by the bare interaction on the SM wave functions.
Formally, one wants to derive an effective operator $\Theta_{\rm eff}$
such that
\begin{equation}
\langle \tilde{\Psi}_{\alpha} | \Theta | \Psi_{\beta} \rangle = \langle
\tilde{\psi}_{\alpha} | \Theta_{\rm eff} | \psi_{\beta} \rangle \label{effop}.  
\end{equation}

The perturbative expansion of effective operators has been approached
since the earliest attempts to employ realistic potentials for SM
calculations, and among many they should be mentioned the fundamental
and pioneering studies carried out by L. Zamick for the problematics of
electromagnetic transitions
\citep{Mavromatis66,Mavromatis67,Federman69} and I. S. Towner for the
study of the quenching of spin-operator matrix elements
\citep{Towner83,Towner87}.

In present section we discuss the formal structure of non-Hermitian
effective operators, as introduced by Suzuki and Okamoto in
Ref. \citep{Suzuki95}.     
More precisely, we provide an expansion formula for the effective
operators in terms of the $\hat{\Theta}$-box, that analogously to the
$\hat{Q}$-box in the effective interaction theory (see Sec,
\ref{perturbative}), is the building block for constructing effective
operators.   

According to Eq. (\ref{defheff}) (and keeping in mind that $\omega
\equiv Q \omega P$), we may write \heff~as
\begin{equation}
H_{\rm eff} = P H (P + \omega),
\end{equation}
\noindent
so that we can express the true eigenstates $|\Psi_\alpha \rangle$ and
their orthonormal counterparts $\langle \tilde{\Psi}_\alpha |$ as
\begin{equation}
|\Psi_\alpha \rangle = (P + \omega) |\psi_\alpha \rangle ~~~~~~~~~~ \langle
\tilde{\Psi}_\alpha | = \langle \tilde{\psi}_\alpha | (P +
\omega^\dagger \omega ) (P + \omega^\dagger) . 
\end{equation}
On the other hand, a general effective operator expression in the
bra-ket representation is given by
\begin{equation}
\Theta_{\rm eff} = \sum_{\alpha \beta} |\psi_\alpha \rangle \langle
\tilde{\Psi}_\alpha | \Theta | \Psi_\beta \rangle \langle
\tilde{\psi}_\beta | ,
\end{equation}
where $\Theta$ is a general time-independent Hermitian
operator. Therefore we can write $\Theta_{\rm eff}$ in an operator
form as
\begin{equation}
\Theta_{\rm eff} = (P + \omega^\dagger \omega )^{-1} (P + \omega^\dagger)
\Theta (P + \omega).
\end{equation}

It is worth noting that Eq. (\ref{effop}) holds independently of the
normalization of $|\Psi_\alpha \rangle$ and $|\psi_\alpha \rangle$,
but if the true eigenvectors are normalized, then $\langle
\tilde{\Psi}_\alpha | = \langle \Psi_\alpha | $  and the $
|\psi_\alpha \rangle $ should be normalized in the following way
\begin{equation}
\langle \tilde{\psi}_\alpha |(P + \omega^\dagger\omega)| \psi_\alpha \rangle = 1 .
\end{equation}

To explicitly calculate $\Theta_{\rm eff}$, we introduce the
$\hat{\Theta}$-box defined as
\begin{equation}
  \hat{\Theta} = (P + \omega^\dagger) \Theta (P + \omega)\label{thetaboxdef},
\end{equation}
\noindent
so that $\Theta_{\rm eff}$ can be factorized as
\begin{equation}
\Theta_{\rm eff} = (P + \omega^\dagger \omega )^{-1} \hat{\Theta} .
\end{equation}
The derivation of $\Theta_{\rm eff}$ is divided in two parts: the
calculation of $\hat{\Theta}$ and the one of $\omega^\dagger \omega $.

According to Eq. (\ref{thetaboxdef}) and taking into account the
expression of $\omega$ in terms of \heff
\begin{equation}
\omega = \sum_{n=0}^\infty (-1)^n \left (\frac{1}{\epsilon_0 - QHQ}
\right )^{n+1}QH_1P(H_1^{\rm eff})^n~,
\end{equation}
we can write 
\begin{equation}
  \hat{\Theta} = \hat{\Theta}_{PP} + (\hat{\Theta}_{PQ} + h.c.) +
  \hat{\Theta}_{QQ}~,
\end{equation}
where
\begin{equation}
  \hat{\Theta}_{PP} = P \Theta P~,
\end{equation}
\begin{equation}
  \hat{\Theta}_{PQ} = P \Theta \omega P = \sum_{n=0}^\infty
  \hat{\Theta}_n (H_1^{\rm eff})^n~,
\end{equation}
\begin{equation}
  \hat{\Theta}_{QQ} = P \omega^\dagger \Theta \omega P =
  \sum_{n,m=0}^\infty (H_1^{\rm eff})^n
  \hat{\Theta}_{nm} (H_1^{\rm eff})^m~,
\end{equation}
\noindent
and $\hat{\Theta}_m$, $\hat{\Theta}_{mn}$ have the following
expressions:
\begin{eqnarray}
\hat{\Theta}_m & = & \frac {1}{m!} \frac {d^m \hat{\Theta}
 (\epsilon)}{d \epsilon^m} \biggl|_{\epsilon=\epsilon_0} ~, \\
\hat{\Theta}_{mn} & = & \frac {1}{m! n!} \frac{d^m}{d \epsilon_1^m}
\frac{d^n}{d \epsilon_2^n}  \hat{\Theta} (\epsilon_1 ;\epsilon_2)
\biggl|_{\epsilon_1= \epsilon_0, \epsilon_2  = \epsilon_0} ~,
\end{eqnarray}

\noindent
with
\begin{eqnarray}
\hat{\Theta} (\epsilon) = & P \Theta P + P \Theta Q
\frac{1}{\epsilon - Q H Q} Q H_1 P ~, ~~~~~~~~~~~~~~~~~~~\label{thetabox} \\
\hat{\Theta} (\epsilon_1 ; \epsilon_2) = & P H_1 Q
\frac{1}{\epsilon_1 - Q H Q} Q \Theta Q \frac{1}{\epsilon_2 - Q H Q} Q
H_1 P ~.~~~~~ 
\end{eqnarray}

As regards the product $\omega^\dagger \omega$, using the definition
(\ref{eqb}), we can write 
\begin{equation}
\omega^\dagger \omega = -\sum_{n=1}^\infty\sum_{m=1}^\infty((H_1^{\rm eff})^\dagger)^{n-1}\hat{Q}(\epsilon_0)_{n+m-1}(H_1^{\rm eff})^{m-1}.
\end{equation}
Expressing $H_1^{\rm eff}$ in in terms of the $\hat{Q}$-box and its
derivatives (see Eqs. (\ref{kkeq},\ref{kkeqexp})), the above quantity may be
rewritten as 
\begin{equation}
\omega^\dagger \omega = - {\hat{Q}_1 + (\hat{Q}_2\hat{Q} + h.c.) +
  (\hat{Q}_3\hat{Q}\hat{Q} + h.c.) + (\hat{Q}_2\hat{Q}_1\hat{Q} +
  h.c.) + \cdots } \label{omega+omega} 
\end{equation}

Putting together Eq. (\ref{thetabox}) and (\ref{omega+omega}), we can write
the final perturbative expansion form of the effective operator
$\Theta_{\rm eff}$
\begin{equation}
\Theta_{\rm eff}  =  (P + \hat{Q}_1 + \hat{Q}_1 \hat{Q}_1 + \hat{Q}_2
\hat{Q} + \hat{Q} \hat{Q}_2 + \cdots) \times (\chi_0+ \chi_1 + \chi_2
+\cdots)~~, \label{effopexp1} 
\end{equation}
where
\begin{eqnarray}
\chi_0 &=& (\hat{\Theta}_0 + h.c.)+ \hat{\Theta}_{00}~~,  \label{chi0} \\
\chi_1 &=& (\hat{\Theta}_1\hat{Q} + h.c.) + (\hat{\Theta}_{01}\hat{Q}
+ h.c.) ~~, \\
\chi_2 &=& (\hat{\Theta}_1\hat{Q}_1 \hat{Q}+ h.c.) +
(\hat{\Theta}_{2}\hat{Q}\hat{Q} + h.c.) +
(\hat{\Theta}_{02}\hat{Q}\hat{Q} + h.c.)+  \hat{Q} 
\hat{\Theta}_{11} \hat{Q}~~. \label{chin} \\
&~~~& \cdots \nonumber
\end{eqnarray}

It is worth enlightening the strong link existing between \heff~and
any effective operator. 
This is achieved by inserting the identity $\hat{Q}  \hat{Q}^{-1} =
\mathbf{1}$ in Eq. (\ref{effopexp1}), so to obtain the following
expression:

\begin{eqnarray}
\Theta_{\rm eff} & = & (P + \hat{Q}_1 + \hat{Q}_1 \hat{Q}_1 + \hat{Q}_2
\hat{Q} + \hat{Q} \hat{Q}_2 + \cdots) \hat{Q}  \hat{Q}^{-1} \times (\chi_0+ \chi_1 + \chi_2 +\cdots) = \nonumber \\
~ & = & H_{\rm eff} \hat{Q}^{-1}  (\chi_0+ \chi_1 + \chi_2 +\cdots) ~~.
\label{effopexp2}
\end{eqnarray}

In actual calculations the $\chi_n$ series is arrested to a finite
order and the starting point is the derivation of a perturbative
expansion of $\hat{\Theta}_0 \equiv \hat{\Theta}(\epsilon_0)$ and
$\hat{\Theta}_{00} \equiv \hat{\Theta}(\epsilon_0;\epsilon_0)$,
including diagrams up to a finite order in the perturbation theory,
consistently with the expansion of the $\hat{Q}$-box. 
The issue of the convergence of the $\chi_n$ series and of the
perturbative expansion of $\hat{\Theta}_0$ and $\hat{\Theta}_{00}$ will
be extensively treated in section \ref{doublebeta}.

\begin{figure}[h!]
\begin{center}
\includegraphics[width=10cm]{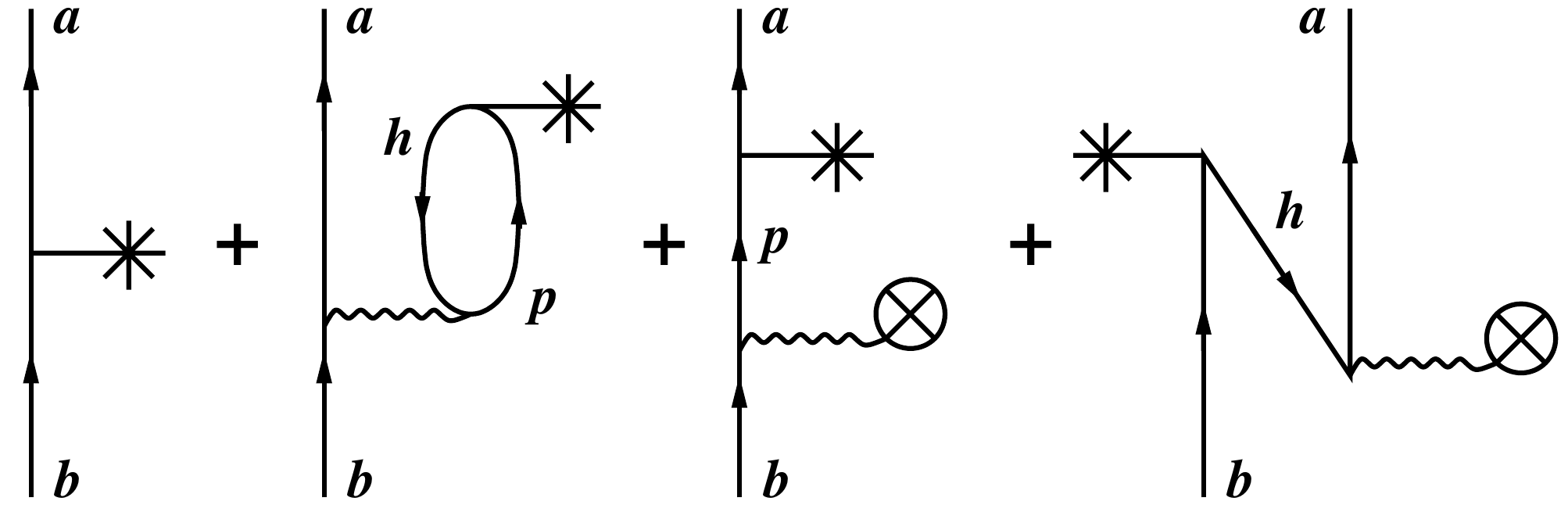}
\caption{One-body second-order diagrams included in the perturbative
  expansion of $\hat{\Theta}_0$. The asterisk indicates the bare operator
  $\Theta$.}
\label{figeffop1}
\end{center}
\end{figure}

In Fig. \ref{figeffop1} we report all the diagrams up to second order
appearing in the $\hat{\Theta}_0$ expansion for a one-body operator
$\Theta$.

The evaluation of the diagrams involved in the derivation of
$\Theta_{\rm eff}$ follows the same rules described in the previous
section.
In the following, therefore, we will just outline the procedure to
calculate such diagrams with one $\Theta$ vertex.

Let us suppose that the operator $\Theta$ transforms like a spherical
tensor of rank $\lambda$, component $\mu$: 
\begin{equation}
\Theta \equiv T^\lambda_\mu~,
\end{equation}
with
\begin{equation}
(T^\lambda_\mu)^\dagger = (-1)^{\lambda - \mu}T^\lambda_{-\mu}~.
\end{equation}

By using the Wigner-Eckart theorem, it is possible to express any
transition matrix element in terms of a reduced transition element
\begin{equation}
\langle j_a || T^\lambda || j_b \rangle = (-1)^{\lambda - \mu} \langle
j_a | T^\lambda_\mu | j_b \rangle ~, \label{redmatop}
\end{equation}
\noindent
where in the r.h.s. of Eq. (\ref{redmatop}) $j_b$ is coupled to $j_a$
to a total angular momentum and projection equal to $\lambda$ and
$-\mu$, respectively, and we have assumed, without any lack of
generality, that we are dealing with single-particle states.

Therefore, we evaluate each diagram as a contribution to the reduced
matrix element of the effective operator.
To be more explicit, we consider as an example the calculation of the
following second-order diagram that takes into account the
renormalization of the operator due to $1p$-$1h$ core excitations.

\begin{figure}[!]
\begin{center}
\includegraphics[width=6cm]{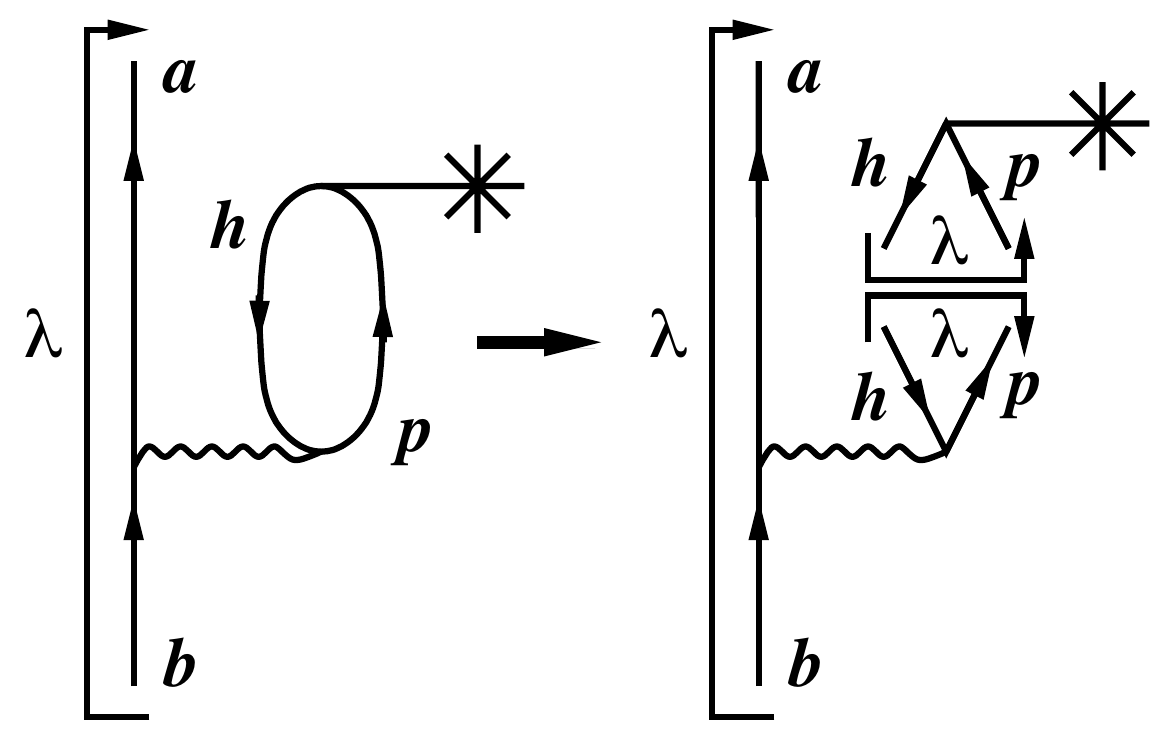}
\caption{One-body second-order 2p-1h diagram included in the perturbative
  expansion of $\hat{\Theta}_0$. The cross indicates the bare operator
  $\Theta$.}
\label{figeffop1b}
\end{center}
\end{figure}

The first step is to couple $j_b$ and $j_a$ to a total angular
momentum equal to $\lambda$.
This enables to factorize the diagram as the product of a
cross-coupled matrix element of the interaction and the reduced matrix
element of the operator (see right-hand side of
Fig. \ref{figeffop1b}).

Explicitly, we can evaluate the diagram as
\begin{equation}
  \langle j_a || \Theta_{2p1h} || j_b \rangle = -
   \sum_{p,h} (-1)^{j_p+j_h-\lambda} \frac{\langle j_a,p; \lambda| V_{NN} |
    j_b,h; \lambda\rangle_{\rm{\bf CC}}
\langle h || T^\lambda || p \rangle
  }
{\epsilon_0 - (\epsilon_a+\epsilon_b-\epsilon_h)}~.
\end{equation}

The minus sign in front of the value is due to the fact that 
$n_h=n_l=1$, and that an extra-phase factor $(-1)^{n_{\rm ph}}$ is
needed for the total number of cut of particle-hole pairs ($n_{\rm
  ph}$) \citep{Kuo81}, since we have cut the inner loop to factorize
the diagram.  
\vspace{0.4truecm}

\section{Applications}\label{applications}
\vspace{0.1truecm}
We present in this section a specific example of SM calculations which
have been performed by employing effective SM Hamiltonian a decay
operators derived from realistic nuclear potentials within the
many-body perturbation theory.

It should be noted that this kind of calculations have been carried
out since the middle of 1960s, but the spirit has been mostly to
retain only the TBMEs, since the single-body component of
\heff~has not been considered enough accurate to provide SM results in
a good agreement with experiment.
A large sample of calculations which are framed in such a successful
approach can be found in previous reviews on this topic
\citep{Hjorth95,Coraggio09a}.

We have sampled in this work the results of a calculation where both
SP energies and TBMEs, that are needed to diagonalize the SM
Hamiltonian, have been obtained by deriving \heff~according to the
procedures that have been reported in the previous section.
Besides \heff, the many-body perturbation theory has been employed to
derive consistently effective operators to calculate electromagnetic
transition rates and Gamow-Teller (GT) strengths without resorting to
empirical effective charges or quenching factors for the axial
coupling constant $g_A$.

There are some motivations that lead to perform SM calculations by
deriving and employing all SM parameters - SP energies, TBMEs,
effective transition and decay operators - starting from realistic
nuclear forces:
\begin{itemize}
\item the need to study the soundness of many-body perturbation theory
  in order to provide reliable SM parameters;
\item to investigate the ability of classes of nuclear potentials
  to describe nuclear structure observables;
\item the opportunity to compare and benchmark SM calculations with
  other nuclear structure methods which employ realistic potentials.
\end{itemize}

The goal of these studies is to test the reliability of such an
approach to the nuclear shell model, especially its predictiveness
that is crucial to describe physical phenomena that are not yet at
hand experimentally.

\vspace{0.4truecm}

\subsection{The double-$\beta$ decay around doubly-closed
  $^{132}$Sn}\label{doublebeta}
\vspace{0.1truecm}
Neutrinoless double-$\beta$ ($0\nu\beta\beta$) decay is an exotic
second-order electroweak process predicted by extensions of the
Standard Model of particle physics.
Observation of such a process would show the non-conservation of
lepton number, and evidence that neutrinos have a Majorana mass
component (see Refs. \citep{Avignone08,Vergados12} and references
therein).

In the framework of light-neutrino exchange, the
half life for the $0\nu\beta\beta$ decay is inversely proportional to
the square of the effective Majorana neutrino mass $\langle m _{\nu}
\rangle $:
\begin{equation}
\left[ T^{0\nu}_{1/2}\right]^{-1} = G^{0\nu} \left| M^{0\nu} \right|^2
g_A^4 \left| \frac{ \langle m_{\nu}\rangle}{m_e} \right|^2~,
\label{halflife}
\end{equation}

\noindent
where $g_A$ is the axial coupling constant, $m_e$ is the electron
mass, $G^{0\nu}$ is the so-called phase-space factor (or kinematic
factor), and $M^{0\nu}$ is the nuclear matrix element (NME), that is
connected to the wave functions of the nuclei involved in the decay.

At present, the phase-space factors for nuclei that are possible
candidates for $0\nu\beta\beta$ decay may be calculated with great
accuracy \citep{Kotila12,Stoica13}.
Therefore, it is crucial to have precise values of the NME, both to
improve the reliability of the $0\nu\beta\beta$ lifetime predictions -
fundamental ingredient to design new experiments - and to extract
neutrino properties from the experimental results, when they will
become available.

Several nuclear structure models are exploited to provide NME values
as precise as possible, the most largely employed being, at present,
the Interacting Boson Model (IBM) \citep{Barea09,Barea12,Barea13}, the
Quasiparticle Random-Phase Approximation (QRPA)
\citep{Simkovic08,Simkovic09,Fang11,Faessler12}, Energy Density
Functional methods \citep{Rodriguez10}, the Covariant Density
Functional Theory \citep{Song14,Yao15,Song17}, the
Generator-Coordinate Method (GCM) \citep{Jiao17,Yao18,Jiao18,Jiao19},
and the Shell Model (SM)
\citep{Menendez09a,Menendez09b,Horoi13b,Neacsu15,Brown15}.

All the above models use a truncated Hilbert space
to overcome the computational complexity, and each of them can be more
efficient than another for a specific class of nuclei.
However, when comparing the calculated NMEs obtained with different
approaches, it can be shown that, at present, the results 
can differ by a factor of two or three (see for instance the review in Ref.
\citep{Engel17}).

In Ref. \citep{Coraggio20a}, it has been reported on the calculation of
the $0\nu\beta\beta$-decay NME for  $^{48}$Ca, $^{76}$Ge,
$^{82}$Se, $^{130}$Te, and $^{136}$Xe in the framework of the realistic SM,
where \heffs~ and $0\nu\beta\beta$-decay effective operators are
consistently derived starting from a realistic $NN$ potential, the
high-precision CD-Bonn potential \citep{Machleidt01b}. 

It is worth mentioning that the above work is not the first example of
such an approach, since it has been pioneered by Kuo
and coworkers \citep{Song89,Staudt92}, and more recently pursued by
Holt and Engel \citep{Holt13d}.

In present work, we will restrict ourselves to present
the results obtained in Ref. \citep{Coraggio20a} for the heavy-mass
nuclei around $^{132}$Sn, $^{130}$Te and $^{136}$Xe. 
At present, these nuclei are under investigation as
$0\nu\beta\beta$-decay candidates by some large experimental collaborations.
The possible $0\nu\beta\beta$ decay of $^{130}$Te is explored by the CUORE
collaboration at the INFN Laboratori Nazionali del Gran Sasso in Italy
\citep{CUORE}, while $^{136}$Xe is investigated both by
the EXO-200 collaboration at the Waste Isolation Pilot Plant in
Carlsbad, New Mexico, \citep{EXO-200}, and by the KamLAND-Zen
collaboration in the Kamioka mine in Japan \citep{Kamland16}.

The starting point of the SM calculation is the high-precision CD-Bonn
$NN$ potential \citep{Machleidt01b}, whose non-perturbative behavior
induced by its repulsive high-momentum components is healed resorting
to the so-called \vlwk~approach \citep{Bogner02}.

This provides a smooth potential which preserves exactly the onshell
properties of the original $NN$ potential up to a chosen cutoff
momentum $\Lambda$.
As in other SM studies
\citep{Coraggio15a,Coraggio15b,Coraggio16a,Coraggio17a}, the value of
the cutoff has been chosen as $\Lambda=2.6$ fm$^{-1}$, since the role
of missing three-nucleon force (3NF) decreases by enlarging the
\vlwk~cutoff \citep{Coraggio15b}.
In fact, in Ref. \citep{Coraggio15b} it has been shown that
\heffs~derived from \vlwks~with small cutoffs ($\Lambda=2.1$
fm$^{-1}$)  own SP energies in a worse agreement with experiment, and
also an unrealistic shell-evolution behavior.
This characteristic may be ascribed to larger impact of the induced
3NF, which becomes less important by enlarging the cutoff.

We have experienced that $\Lambda=2.6$ fm$^{-1}$, within a
perturbative expansion of the \qbox, is an upper limit, since a larger
cutoff worsens the order-by-order behavior of the perturbative
expansion, and at the end of present section it is reported a study of
the perturbative properties of \heff~and of the effective decay
operators derived using this \vlwk~potential.

The Coulomb potential is explicitly taken into account in the
proton-proton channel.

The shell-model effective hamiltonian $H_{\rm eff}$ has been derived
within the framework of the many-body perturbation theory as described
in section \ref{perturbative}, including diagrams up to third order in $H_1$ in
the \qbox-expansion, while all the effective operators, both one- and
two-body, have been obtained consistently using the approach described
in section \ref{effopsec}, including diagrams up to third order in
perturbation theory in the evaluation of the $\hat{\Theta}$-box and
arresting the $\chi_n$ series in Eq. \eqref{effopexp1} to $\chi_2$.  

The effective hamiltonian and operators are defined in a model space
spanned by the five $0g_{7/2},1d_{5/2},1d_{3/2},2s_{1/2},0h_{11/2}$
proton and neutron orbitals outside the doubly-closed $^{100}$Sn
core. 
The single-particle (SP) energies, and the two-body matrix elements
(TBMEs) of \heff~ can be found in Ref. \citep{Coraggio17a}.

Before showing the results for the $0\nu\beta\beta$ NME obtained in
Ref. \citep{Coraggio20a}, it is worth checking the reliability of the
adopted approach.
To this end, we present here some results obtained for the calculation
of quantities for which there exist an experimental conterpart to
compare with. Namely, we show some selected results for the
electromagnetic properties, GT strength distributions and
$2\nu\beta\beta$ decays in $^{130}$Te and $^{136}$Xe, that are
reported in Refs. \citep{Coraggio17a,Coraggio19a}. 

\begin{figure}[h!]
\begin{center}
  \includegraphics[scale=0.40,angle=0]{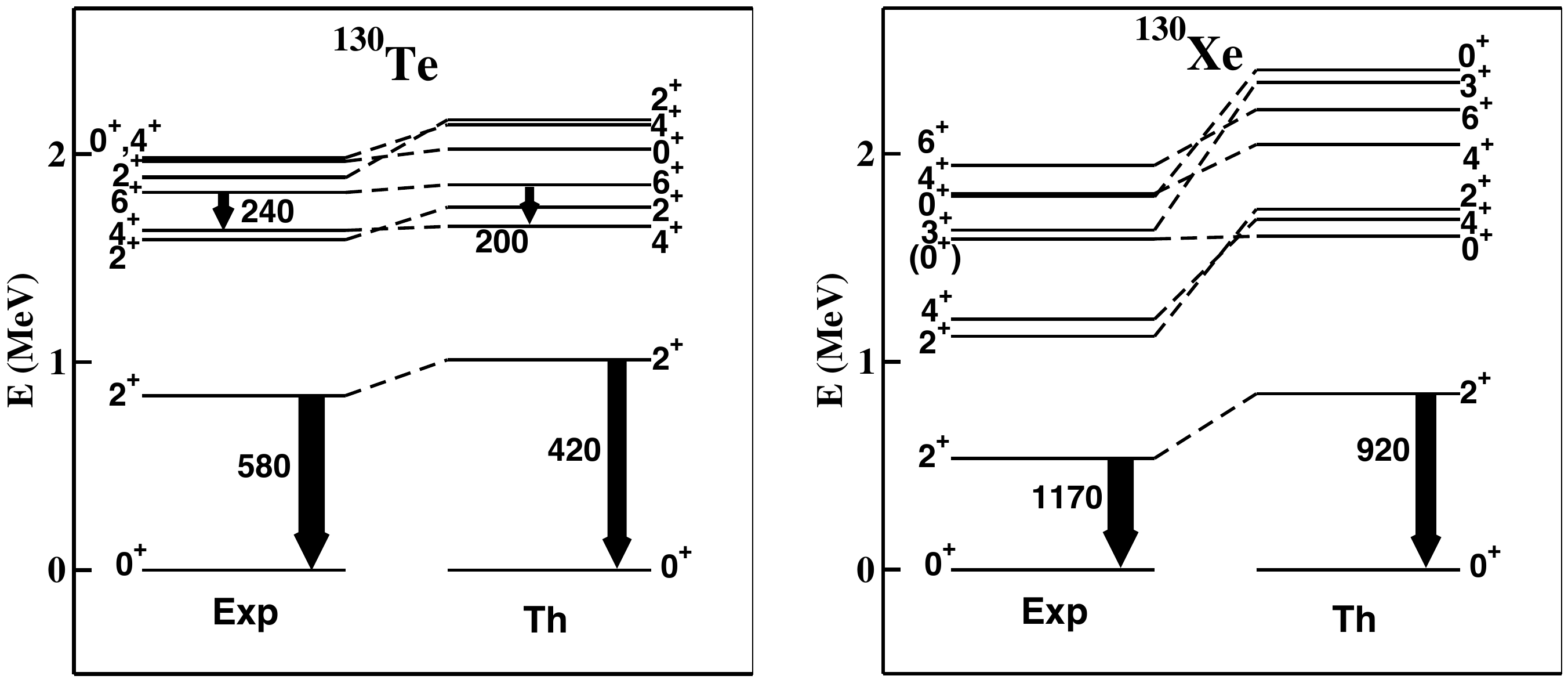}
\caption{Experimental and calculated spectra of $^{130}$Te and
  $^{130}$Xe. The arrows are proportional to the $B(E2)$ strengths,
  whose values are reported in $e^2{\rm fm}^4$. Reproduced from
  Ref. \citep{Coraggio19a}.} 
\label{130Te130Xe}
\end{center}
\end{figure}

\begin{figure}[ht]
  \begin{center}
    \includegraphics[scale=0.40,angle=0]{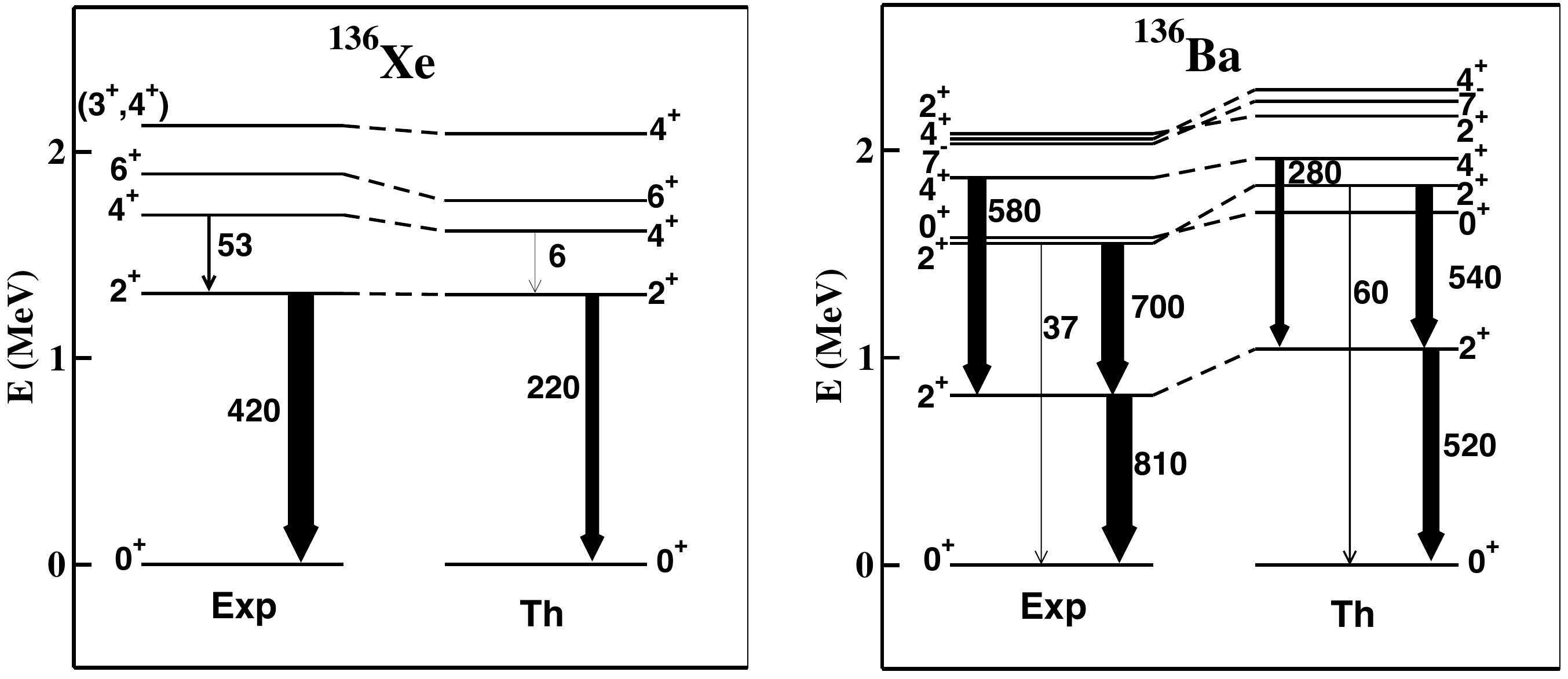}
\caption{ Same as in Fig. \ref{130Te130Xe}, but for $^{136}$Xe and
  $^{136}$Ba. Reproduced from
  Ref. \citep{Coraggio19a}.}
\label{136Xe136Ba}
  \end{center}
\end{figure}

In Figs. \ref{130Te130Xe} and \ref{136Xe136Ba} they are shown the
experimental \citep{ensdf,xundl} and calculated low-energy spectra and
$B(E2)$s of parent and grand-daughter nuclei involved in
double-beta-decay of $^{130}$Te and $^{136}$Xe, respectively. 

From the inspection of Figs. \ref{130Te130Xe} and \ref{136Xe136Ba}, it
can be seen that the comparison between theory and experiment, as
regards the low-lying excited states and the $B(E2)$ transition rates,
is quite good for $^{130}$Te, $^{136}$Xe, and $^{136}$Ba, while it is
less satisfactory for $^{130}$Xe, whose theoretical spectrum is
expanded when compared with the observed one.
As regards the electromagnetic properties, in Ref. \citep{Coraggio19a}
they are calculated also some $B(M1)$ strengths and
magnetic dipole moments using an effective spin-dependent $M1$
operator, and the comparison with the available data (see Tables VII and
IX in Ref. \citep{Coraggio19a}) evidences a good agreement.

Two different kinds of experimental data related to GT decay are available in
$^{130}$Te and $^{136}$Xe: GT-strength distributions, and the NMEs
involved in $2\nu\beta\beta$ decays.

The GT strength $B({\rm GT})$ can be extracted from the GT component
of the cross section at zero degree of intermediate energy charge-exchange
reactions, following the standard approach in the distorted-wave Born
approximation (DWBA) \citep{Goodman80,Taddeucci87}:

\begin{equation}
\frac{d\sigma^{GT}(0^\circ)}{d\Omega} = \left (\frac{\mu}{\pi \hbar^2} \right
)^2 \frac{k_f}{k_i} N^{\sigma \tau}_{D}| J_{\sigma \tau} |^2 B(GT)~~,
\end{equation}

\noindent
where $N^{\sigma \tau}_{D}$ is the distortion factor, $| J_{\sigma
  \tau} |$ is the volume integral of the effective $NN$ interaction,
$k_i$ and $k_f$ are the initial and final momenta, respectively, and
$\mu$ is the reduced mass.

On the other hand, the experimental $2\nu\beta\beta$ NME $M_{\rm
  GT}^{2\nu}$ can be extracted from the observed half life
$T^{2\nu}_{1/2}$ of the parent nucleus 

\begin{equation}
\left[ T^{2\nu}_{1/2} \right]^{-1} = G^{2\nu} \left| M_{\rm GT}^{2\nu}
\right|^2 ~~.
\label{2nihalflife}
\end{equation}

Both the above quantities can be calculated in terms of the matrix
elements of the GT$^-$ operator $\vec{\sigma} \tau^-$
\begin{equation}
B(GT) = \frac{ \left| \langle \Phi_f || \sum_{j}
  \vec{\sigma}_j \tau^-_j || \Phi_i \rangle \right|^2} {2J_i+1}~~,
\label{GTstrength}
\end{equation}
\begin{equation}
M_{\rm GT}^{2\nu}= \sum_n \frac{ \langle 0^+_f || \vec{\sigma} \tau^-
  || 1^+_n \rangle \langle 1^+_n || \vec{\sigma}
\tau^- || 0^+_i \rangle } {E_n + E_0} ~~,
\label{doublebetame}
\end{equation}

\noindent
where $E_n$ is the excitation energy of the $J^{\pi}=1^+_n$
intermediate state, $E_0=\frac{1}{2}Q_{\beta\beta}(0^+) +\Delta M$,
$Q_{\beta\beta}(0^+)$ and $\Delta M$ being the $Q$ value of the $\beta
\beta$ decay and the mass difference between the daughter and parent
nuclei, respectively.
The nuclear matrix elements in Eqs. \eqref{GTstrength} and
\eqref{doublebetame} are calculated within the long-wavelength
approximation, including only the leading order of the Gamow-Teller
operator in a nonrelativistic reduction of the hadronic current.

In Ref. \citep{Coraggio19a}, the GT strength distributions and the
$2\nu\beta\beta$ NMEs have been calculated for $^{130}$Te and
$^{136}$Xe using an effective spin-isospin dependent GT operator,
derived consistently with \heff~by following the procedure described
in section \ref{effopsec}.

\begin{figure}[h!]
\begin{minipage}{20pc}
\begin{center}
\includegraphics[width=20pc]{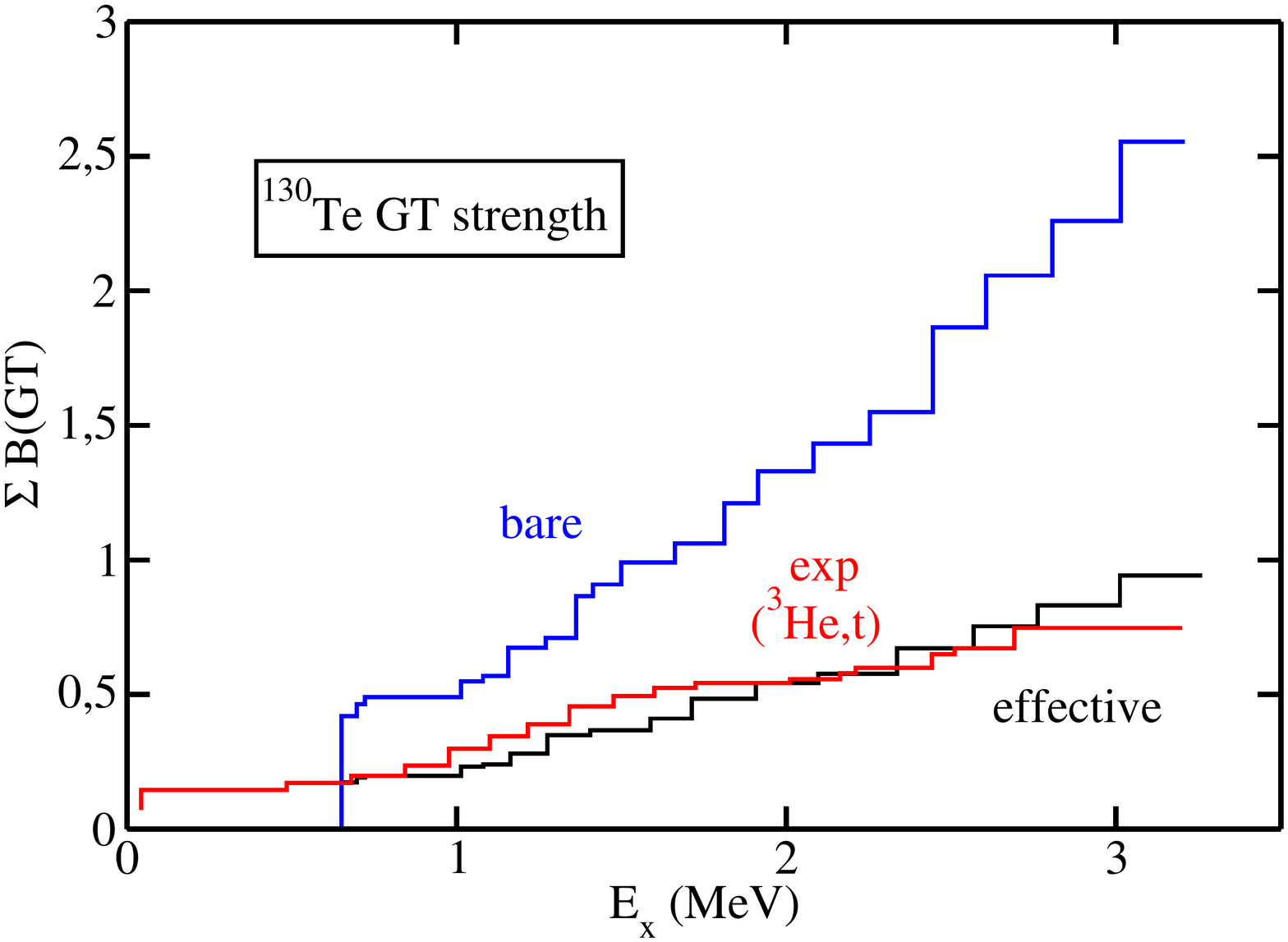}
\end{center}
\end{minipage}
\begin{minipage}{2pc}
$~~$
\end{minipage}
\begin{minipage}{20pc}
\begin{center}
\includegraphics[width=20pc]{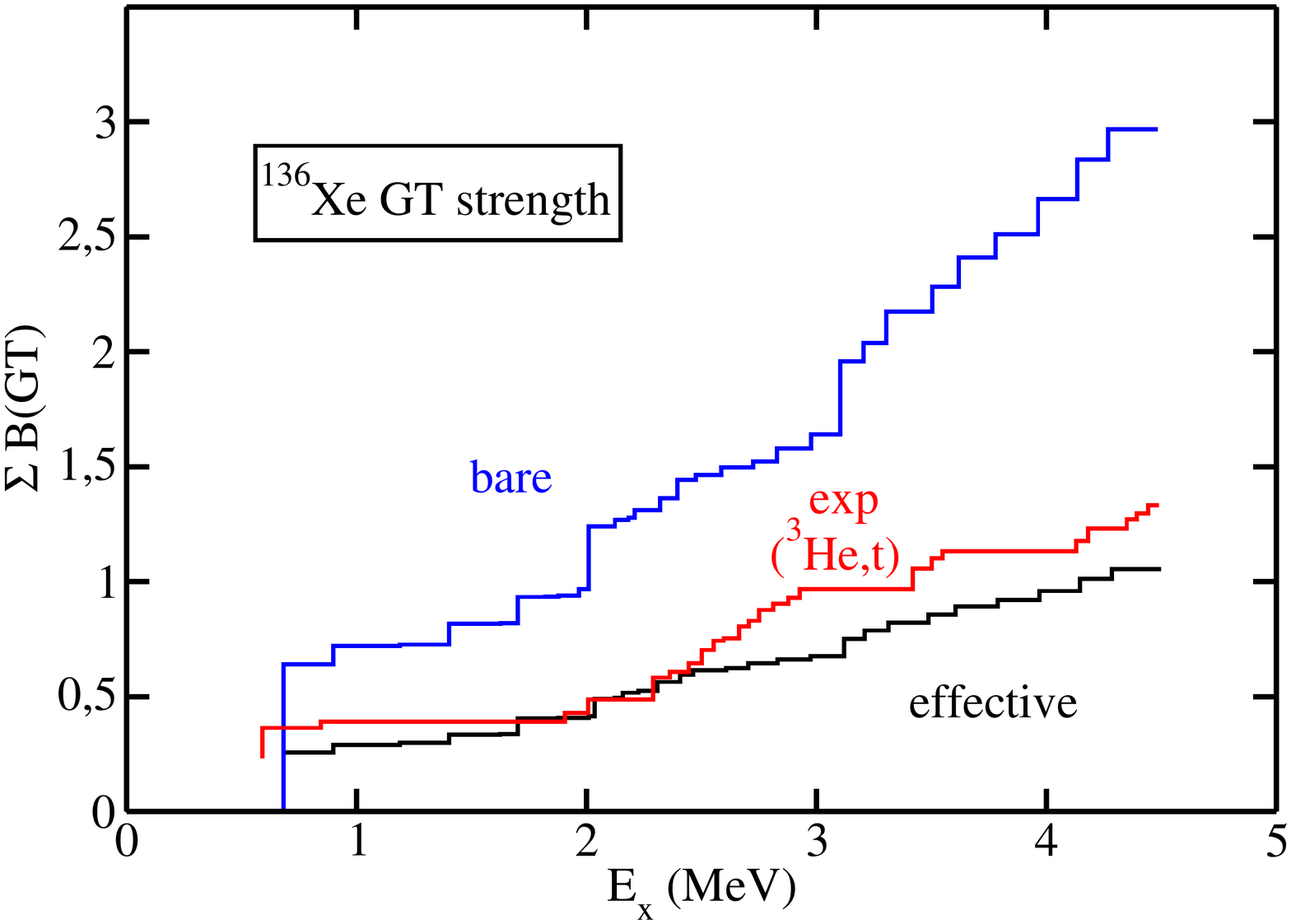}
\end{center}
\end{minipage} 
\caption{Running sums of the $B(GT)$
  strengths as a function of the excitation energy $E_x$ up to 3 MeV,
  and 4.5 MeV respectively for $^{130}$Te and $^{136}$Xe.
Reproduced from Ref. \citep{Coraggio19a}.}\label{130Te136XeGT-}
\end{figure}

Fig. \ref{130Te136XeGT-} shows the theoretical running
  sums of the GT strengths $\Sigma B({\rm GT})$, calculated  with both
  bare and effective GT operators, reported as a function of the
  excitation energy, and compared with the available data extracted
  from ($^3$He, $t$) charge-exchange experiments
  \citep{Puppe12,Frekers13}, for $^{130}$Te and $^{136}$Xe.
From its inspection, it can be seen that,
in both nuclei, the GT-strength distributions calculated using the
bare GT operator overestimate the experimental ones by more than a
factor of two. 
Including the many-body renormalization of the GT operator brings the
predicted GT strength distribution into much better agreement with
that extracted from experimental data.

In Ref. \citep{Coraggio19a} the NMEs $M_{\rm GT}^{2\nu}$ involved in the decay of
$^{130}$Te and $^{136}$Xe are calculated using the definition in
Eq. \eqref{doublebetame}, by means of the Lanczos strength-function
method as in Ref. \citep{Caurier05}. The results obtained with the bare
GT operator and with the effective one are reported in Table
\ref{ME_130Te136Xe}, and compared with the experimental values
\citep{Barabash15}.  

\begin{table}[ht]
  \caption{\label{ME_130Te136Xe}Experimental \citep{Barabash15}  and
    calculated NME of the $2\nu\beta\beta$ decay (in MeV$^{-1}$) for
    $^{130}$Te and $^{136}$Xe.} 
  \begin{center}
    \begin{tabular}{cccc}
      \hline
      ~ & ~ & ~ & ~ \\
 Decay & NME$_{\rm Expt}$ & bare & effective  \\
\hline
~ & ~ & ~ & ~ \\
$^{130}$Te  $\rightarrow$ $^{130}$Xe & $ 0.031 \pm 0.004$ & 0.131 & 0.061 \\
~ & ~ & ~ & ~ \\
$^{136}$Xe  $\rightarrow$ $^{136}$Ba & $ 0.0181 \pm 0.0007$ & 0.0910 & 0.0341\\
                                                                     
\hline
\end{tabular}
\end{center}
    \end{table}

The effective operator induces a relevant quenching of the calculated
NME, 47\% for $^{130}$Te and 37\% for $^{136}$Xe decay, leading to a
quite good agreement with the experimental value in both nuclei,
that is of the same quality as that of other shell-model calculations
where all parameters (SP energies and TBMEs) have been fitted to
experiment and a quenching factor q has been introduced to
reproduce GT data (see for example Ref. \citep{Caurier12}).
The overall agreement between theory and experiment shows the present
capability of the many-body perturbation theory to derive
consistently effective Hamiltonians and transition operators that are
able to reproduce quantitatively the observed spectroscopic and decay
properties, without resorting to any empirical adjustment, such as the
quenching of the axial coupling constant $g_A$.
This supports the reliability of this approach to
calculate the NME involved in $0\nu\beta\beta$, whose results have
been reported in Ref. \citep{Coraggio20a}, and are shortly recollected
in the following.  

The $0\nu\beta\beta$ two-body operator for the
light-neutrino scenario can be expressed in the closure approximation
(see for instance Refs. \citep{Haxton84,Tomoda91}) in terms of the
neutrino potentials $H_{\alpha}$ and form functions $h_{\alpha}(q)$
($\alpha$ = $F$, GT, or $T$) as follows

\begin{eqnarray} 
 \Theta_{\rm GT} & = & \vec{\sigma}_1 \cdot \vec{\sigma}_2 H_{\rm
GT}(r)\tau^-_{1} \tau^-_{2} \label{operatorGT} \\
\Theta_{\rm F} & = & H_{\rm F}(r)\tau^-_{1} \tau^-_{2}  \label{operatorF} \\ 
\Theta_{\rm T} & = & \left[3\left(\vec{\sigma}_1 \cdot \hat{r} \right) 
\left(\vec{\sigma}_1 \cdot \hat{r} \right) - \vec{\sigma}_1 \cdot
                      \vec{\sigma}_2 \right] H_{\rm T}(r)\tau^-_{1}
\tau^-_{2}~, \label{operatorT} 
\end{eqnarray}

\begin{equation}
H_{\alpha}(r)=\frac {2R}{\pi} \int_{0}^{\infty} \frac {j_{n_{\alpha}}(qr)
  h_{\alpha}(q^2)qdq}{q+\left< E \right>}~.
\label{neutpot}
\end{equation}

The value of the parameter $R$ is $R=1.2~A^{1/3}$ fm, the
$j_{n_{\alpha}}(qr)$ are the spherical Bessel functions,
$n_{\alpha}=0$ for Fermi and Gamow-Teller components, $n_{\alpha}=2$
for the tensor one.
The explicit expression of neutrino form functions $h_{\alpha}(q)$ may
be found in Ref. \citep{Coraggio20a}, and the average energies $\left<
E \right>$ are evaluated as in Refs. \citep{Haxton84,Tomoda91}. 

Apart from effects related to sub-nucleonic degrees of freedom, that
have not been taken into account in Ref. \citep{Coraggio20a}, the
$0\nu\beta\beta$-decay operator has to be renormalized to take into
account both the degrees of freedom that are neglected in the adopted
model space and the contribution of the short-range correlations
(SRC).
The latter arise since the action of a two-body decay operator on an
unperturbed (uncorrelated) wave function, as the one used in the
perturbative expansion of $\Theta_{\rm eff}$, differs from the
action of the same operator on the real (correlated) nuclear wave function.

It is worth pointing out that the calculations for \dbb~
  decay are not affected by this renormalization, since, as mentioned
  before, we retain only the leading order of the long-wavelength
  approximation which corresponds to a zero-momentum-exchange ($q =
  0$) process. 
On the other hand, the inclusion of higher-order contributions or
corrections due to the sub-nucleonic structure of the nucleons
\cite{Park93,Pastore09,Piarulli13,Pastore18} would connect high- and
low-momentum configurations and this renormalization should be carried
out for the two-neutrino emission decay too.

In Ref. \citep{Coraggio19b} the inclusion of SRC has been realized by
means of an original approach \citep{Coraggio19b} that is consistent
with the \vlwk~procedure. 
The \zbb~ operator $\Theta$, expressed in the
momentum space, is renormalized by way of the same similarity
transformation operator $\Omega_{{\rm low}\mbox{-}k}$ that defines the
\vlwk~potential. This enables to consider effectively the
high-momentum (short range) components of the $NN$ potential, in a
framework where their direct contribution is not explicitly considered
above a cutoff $\Lambda$. 
The resulting $\Theta_{{\rm low}\mbox{-}k}$ vertices are then
employed in the perturbative expansion of the $\hat{\Theta}$-box to
calculate $\Theta_{\rm eff}$ using Eq. \eqref{effopexp2}.
More precisely, the perturbative expansion has considered diagrams up to
third order in perturbation theory, including the ones related to the
so-called Pauli blocking effect (see Fig. 2 in
Ref. \citep{Coraggio20a}), and the $\chi_n$ series has been arrested to
$\chi_2$.

In Ref. \citep{Coraggio20a} the contribution of the tensor component of
the neutrino potential (Eq. \eqref{operatorT}) is neglected, and
therefore the total nuclear matrix element $M^{0\nu}$ is expressed as

\begin{equation}
 M^{0 \nu} = M^{0 \nu}_{\rm GT}- \left( \frac{g_{V}}{g_{A}} \right)^2  M^{0
   \nu}_{\rm F}~,
\label{nme00nu}
\end{equation}
\noindent
where $g_A=1.2723, ~ g_V=1$ \citep{PDG18}, and the matrix elements
between the initial and final states $M_\alpha^{0\nu}$ are calculated within the
closure approximation

\begin{equation}
M_\alpha^{0\nu} =   \sum_{j_n j_{n^\prime} j_p j_{p^\prime}}
\langle f | a^{\dagger}_{p}a_{n} a^{\dagger}_{p^\prime} a_{n^\prime} 
| i \rangle  \times  \left< j_p  j_{p^\prime} \mid \Theta_\alpha \mid  j_n j_{n^\prime}
       \right>~. \label{M0nuapp}
\end{equation}

The calculated NMEs using the \zbb-decay effective operator are
reported in Table \ref{NME} and compared with the values obtained with
the bare operator without any renormalization.

\begin{table}[ht]
  \caption{\label{NME}Calculated values of \nme~for $^{130}$Te and $^{136}$Xe
    decay. The first column corresponds to the results 
    ontained employing the bare \zbb-decay operator, the second one to
  the calculations performed with $\Theta_{\rm eff}$.}
  \begin{center}
    \begin{tabular}{ccc}
\hline
 Decay & bare operator & $\Theta_{\rm eff}$ \\
\hline
~ & ~ & ~ \\
$^{130}$Te $\rightarrow$ $^{130}$Xe & 3.27 & 3.16 \\
$^{136}$Xe  $\rightarrow$ $^{136}$Ba & 2.47 & 2.39 \\
\hline
\end{tabular}      
\end{center}
\end{table}

The most striking feature that can be inferred from the inspection of
Table \ref{NME} is that the effects of the renormalization of the \zbb-decay
operator are far less relevant than those observed in the
$2\nu\beta\beta$-decay sector. 

A long standing issue related with the calculation of $M^{0\nu}$ is
the possible interplay between the derivation of the effective
one-body GT operator and the renormalization of the two-body GT
component of the \zbb~operator, some authors presuming that the same
empirical quenching introduced to reproduce the observed GT-decay
properties (single-$\beta$ decay strengths, \nmeds, etc.) should be
also employed to calculate $M^{0\nu}$ (see for instance
Refs. \citep{Suhonen17a,Suhonen17b}). 
Actually, the comparison of the results in Tables
\ref{ME_130Te136Xe} and \ref{NME} shows that the mechanisms
which rule the microscopic derivation of the one-body single-$\beta$
and the two-body \zbb~decay effective operators lead to a considerably
different renormalization, at variance with the above hypothesis.

SM calculations of this section have been performed by employing, as
interaction vertices of the perturbative expansion of the \qbox, a
realistic potential derived from the high-precision CD-Bonn $NN$
potential \citep{Machleidt01b}.
This potential is characterized by a strong repulsive behavior in the
high-momentum regime, so, as mentioned before, it has been
renormalized by deriving a low-momentum $NN$ potential using the
\vlwk~approach \citep{Bogner02}.

As in other SM studies
\citep{Coraggio15a,Coraggio15b,Coraggio16a,Coraggio17a}, the value of
the cutoff has been chosen as $\Lambda=2.6$ fm$^{-1}$, since the role
of missing three-nucleon force (3NF) decreases by enlarging the
\vlwk~cutoff \citep{Coraggio15b}.
This value, within a perturbative expansion of the \qbox, is an upper
limit, since a larger cutoff worsens the order-by-order behavior of
the perturbative expansion.
Here, we report some considerations about the properties of the
perturbative  expansion of \heff~and the SM effective transition
operator, when this ``hard'' \vlwk~is employed to derive SM
Hamiltonian and operators.

Studies of the perturbative properties of the SP energy spacings and
TBMEs are reported in Refs. \citep{Coraggio15b} and
\citep{Coraggio09d}, where \heff~has been derived within the model
space outside $^{132}$Sn and starting from the ``hard'' \vlwk.
However, in Ref. \citep{Coraggio18b} it can be found a systematic
investigation of the convergence properties of theoretical SP energy spectra,
TBMEs, and $2\nu\beta\beta$ NMEs as a function both of the dimension of
the intermediate-state space and the order of the perturbative
expansion.
Moreover, in Ref. \citep{Coraggio20a} they are also reported the
convergence properties of the perturbative expansion of the effective \zbb-decay
operator with respect to the number of intermediate states, and the
truncation both of the order of $\chi_n$ operators and the
perturbative order of the diagrams.

Here, we sketch out briefly these results in order to assess the
reliability of realistic SM calculation performed starting from a
``hard'' \vlwk.

The model space employed for the SM calculations reported in
Ref. \citep{Coraggio18b} is spanned by the  five proton and neutron
orbitals $0g_{7/2},1d_{5/2},1d_{3/2},2s_{1/2},0h_{11/2}$ outside the
doubly-closed $^{100}$Sn, in order to study the \dbb~decay of
$^{130}$Te and $^{136}$Xe.

\begin{figure}[h!]
\begin{center}
\includegraphics[width=16cm]{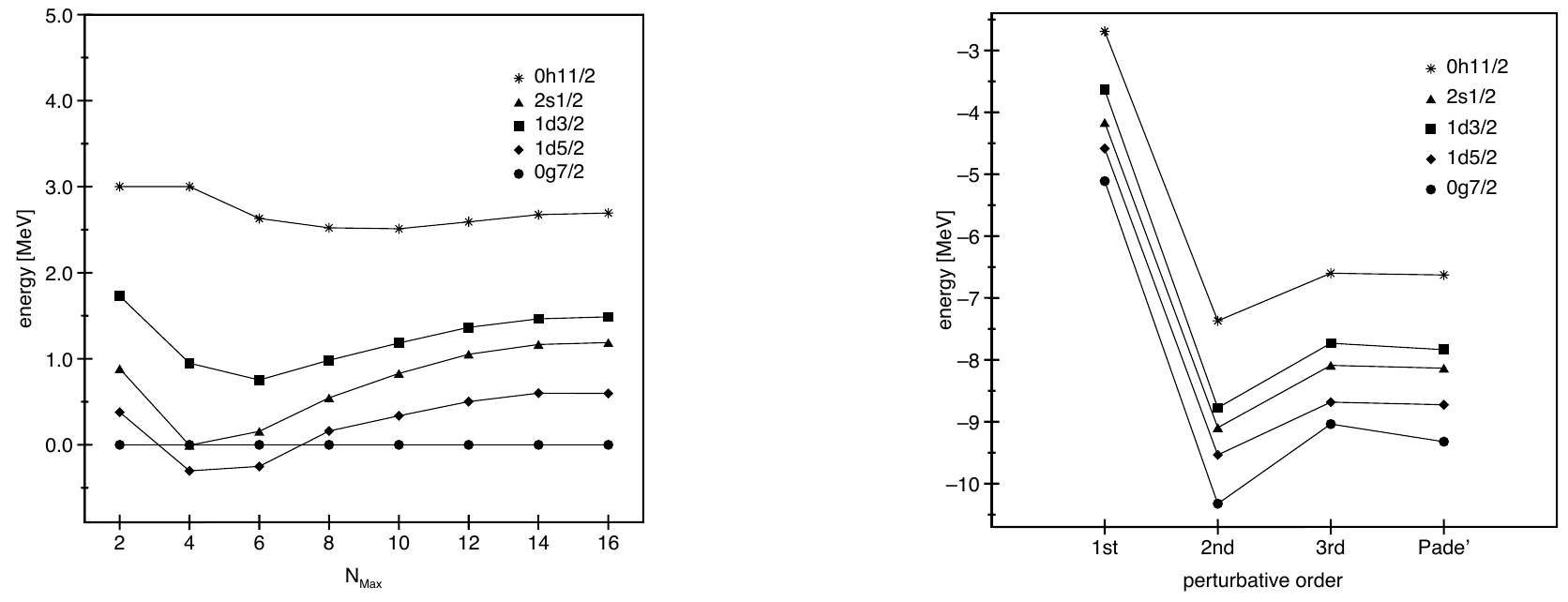}
\caption{\label{convsp}Neutron SP energies as a function of
  $N_{\rm max}$ (left-hand side) of the perturbative order (right-hand
  side). Reproduced from Ref. \citep{Coraggio18b} under the Creative
  Commons CC BY license.} 
\end{center}
\end{figure}

In Fig. \ref{convsp}, it can be found the behavior of the
calculated SP spectrum of $^{101}$Sn, with respect to the $0g_{7/2}$
SP energy, as a function of the maximum allowed excitation energy of
the intermediate states expressed in terms of the oscillator quanta
$N_{\rm max}$.

From the inspection of Fig. \ref{convsp}, it is clear that the
results achieve convergence at $N_{\rm max}=14$, which justifies the
choice, for the perturbative expansion of the effective SM Hamiltonian
and decay operators, to include intermediate states with an
unperturbed excitation energy up to $E_{max}=N_{max} \hbar \omega$
with $N_{max}=16$
\citep{Coraggio17a,Coraggio18b,Coraggio19a,Coraggio20a}.

As regards the order-by-order convergence of the SP energies, in
Fig. \ref{convsp}, the calculated neutron SP energies, using a
number of intermediate states corresponding to $N_{\rm max}=16$, are 
reported as a function of the order of the perturbative expansion up
to the third order.
They are also and compared with the Pad\'e approximant $[2|1]$ of the
$\hat{Q}$-box, which estimates the value to which the perturbative
series may converge.
The results at third order are very close to those
obtained with the Pad\'e approximant, indicating that the truncation
at third order should be a reasonable estimate of the sum of the
series.

\begin{figure}[h!]
\begin{center}
\includegraphics[width=16cm]{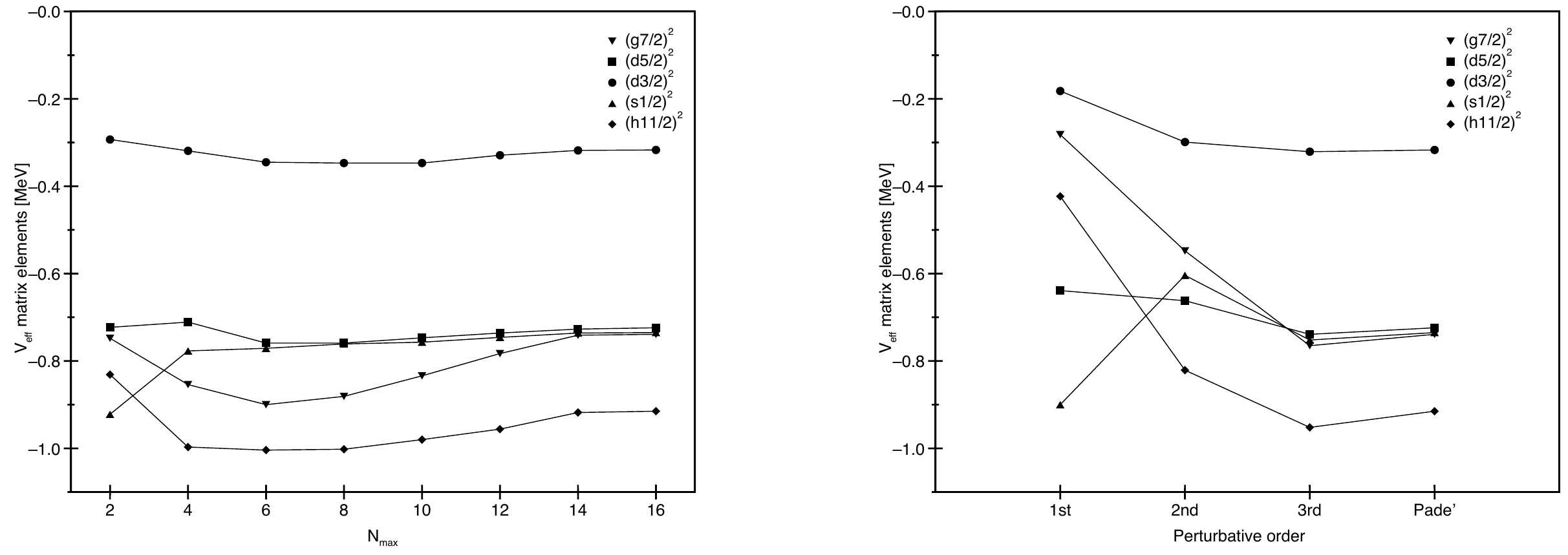}
\caption{\label{convtbme} Neutron-neutron diagonal $J^{\pi}=0^+$
  TBMEs as a function of $N_{\rm max}$ (left-hand side) of the
  perturbative order (right-hand side). Reproduced from
  Ref. \citep{Coraggio18b} under the Creative Commons CC BY license.}
\end{center}
\end{figure}

As regards the TBMEs, we report in Fig. \ref{convtbme} and the
neutron-neutron diagonal $J^{\pi}=0^+$ TBMEs as a function both of
$N_{\rm max}$ and of the perturbative order.
These TBMEs, which contain the pairing properties of the effective
Hamiltonian, are the largest in size of the calculated matrix elements
and the most sensitive to the behavior of the perturbative expansion.

From the inspection of Fig. \ref{convtbme} the convergence with
respect to $N_{\rm max}$ appears to be very fast for diagonal matrix
elements $(1d_{5/2})^2$, $(1d_{3/2})^2$, and $(2s_{1/2})^2$, while
those corresponding to orbitals lacking of their own spin-orbit
partner, $(0g_{7/2})^2$ and $(0h_{11/2})^2$, show a slower convergence.
The order-by-order convergence, as shown in Fig. \ref{convtbme}, is
quite satisfactory, and again the results at third order are very
close to those obtained with the Pad\'e approximant.
Therefore, the conclusion is that \heff, calculated from a
\vlwk~with a cutoff equal to 2.6 fm$^{-1}$ by way of a perturbative
expansion arrested at third order, is a good estimate of the sum of
its perturbative expansion, both for the one- and two-body
components.

Now, we focus the attention on the perturbative expansion of the GT
effective operator ${\rm GT}_{\rm eff}$.

The selection rules of the GT operator, that characterize a
spin-isospin-dependent decay, drive a fast convergence of the matrix
elements of its SM effective operator with respect to $N_{\rm max}$.
In fact, if the perturbative expansion is arrested at second order,
their values do not change from $N_{\rm max}=2$ on \citep{Towner83},
and at third order in perturbation theory their third decimal digit
values do not change from $N_{\rm max}=12$ on.

In Table \ref{nmeobo1} the results of the calculated NME of the
$2\nu\beta\beta$ decays $^{130}{\rm Te}_{\rm g.s.} \rightarrow
^{130}$Xe$_{\rm g.s.}$ and $^{136}{\rm Xe}_{\rm g.s.} \rightarrow
^{136}$Ba$_{\rm g.s.}$, obtained with effective operators at first,
second, and third order in perturbation theory ($\chi_n$ series in
Eq. \eqref{effopexp2} is arrested to $\chi_0$), are reported and
compared with the experimental results \citep{Barabash15}.  

\begin{table}[h]
\caption{\label{nmeobo1} Order-by-order \nmeds~(in MeV$^{-1}$) for
  $^{130}$Te and $^{136}$Xe \citep{Coraggio18b}.}
\begin{center}
\begin{tabular}{ccccc}
  \hline
  &&\\
  Decay & 1st ord \nmed & 2nd ord \nmed &  3rd ord \nmed & Expt. \\
  &&\\
\hline
  &&\\
  $^{130}$Te $\rightarrow$ $^{130}$Xe  & 0.142 & 0.040 & 0.044 & $0.031\pm0.004$ \\ 
  $^{136}$Xe $\rightarrow$ $^{136}$Ba  & 0.0975 & 0.0272 & 0.0285 & $0.0181\pm0.0007$ \\ 
  &&\\
\hline
\end{tabular}
\end{center}
\end{table}

As can be seen, also the order-by-order convergence of the \nmeds~is
very satisfactory, since the results change for both transitions about 260$\%$
from the first- to second-order calculations, while the change is
9$\%$ and 5$\%$ from the second- to third-order results for $^{130}$Te
and $^{136}$Xe decays, respectively.
This suppression of the third-order with respect to the second-order
contribution is favoured by the mutual cancelation of third-order
diagrams.

In Ref. \citep{Coraggio20a} it has been performed also a study about
the convergence properties of the effective decay-operator
$\Theta_{\rm eff}$ for the \zbb~decay, with respect the truncation of
the $\chi_n$ operators, the number of intermediate states which have
been accounted for the perturbative expansion, as well as the
order-by-order behavior up to third order in perturbation theory.

\begin{figure}[h!]
\begin{center}
\includegraphics[width=9cm]{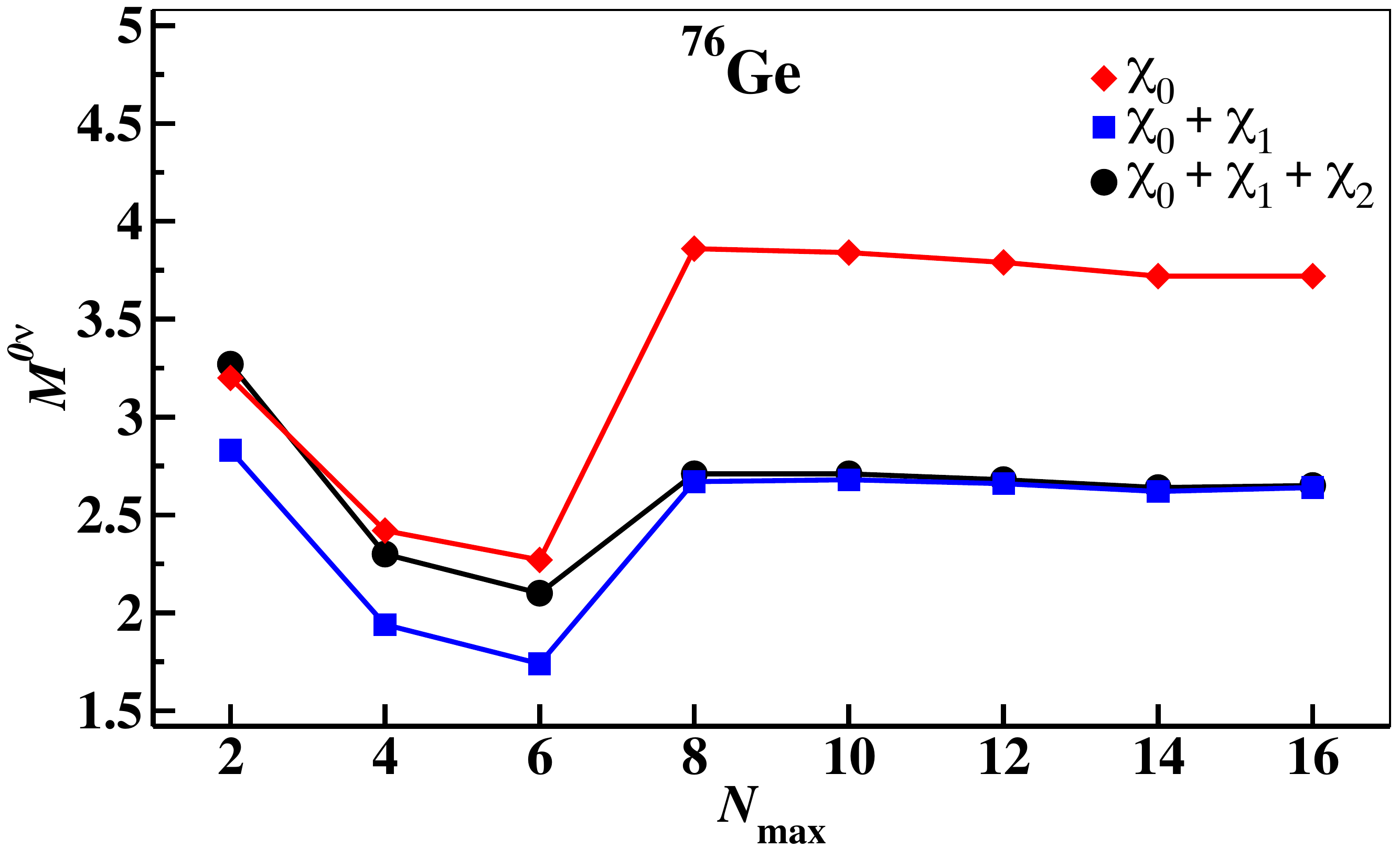}
\caption{\nme~for the $^{76}\mbox{Ge} \rightarrow ^{76}$Se decay as a
  function of $N_{\rm max}$. The red diamonds correspond to a
  truncation of $\chi_n$ expansion up to $\chi_0$, blue squares up to
  $\chi_1$, and black dots up to $\chi_2$. Reproduced from
  Ref. \citep{Coraggio20a}.} 
\label{figNmax}
\end{center}
\end{figure}

In Fig. \ref{figNmax} the results of the calculated values of \nme~for
the ${\rm ^{76}Ge} \rightarrow {\rm ^{76}Se}$ decay are drawn as a function
of the maximum allowed excitation energy of the intermediate states
expressed in terms of the oscillator quanta $N_{\rm max}$, and they
are reported including $\chi_n$ contributions up to $n=2$.
We can see that the \nmes~values are convergent from $N_{\rm max}=12$
on and that contributions from  $\chi_1$ are quite relevant, those
from $\chi_2$ being almost negligible.

It is worth pointing out that, according to expressions \eqref{chin},
$\chi_3$ is defined in terms of the first, second, and third
derivatives of $\hat{\Theta}_0$ and $\hat{\Theta}_{00}$, as well as on
the first and second derivatives of the \qbox.
This means that one could estimate $\chi_3$ being about
one order of magnitude smaller than $\chi_2$ contribution.

On the above grounds, in Ref. \citep{Coraggio20a} the effective SM
\zbb-decay operator has been obtained including in the perturbative
expansion up to third-order diagrams, whose number of intermediate
states corresponds to oscillator quanta up to $N_{\rm  max}=14$, and
up to $\chi_2$ contributions.

\begin{figure}[h!]
\begin{center}
\includegraphics[width=16cm]{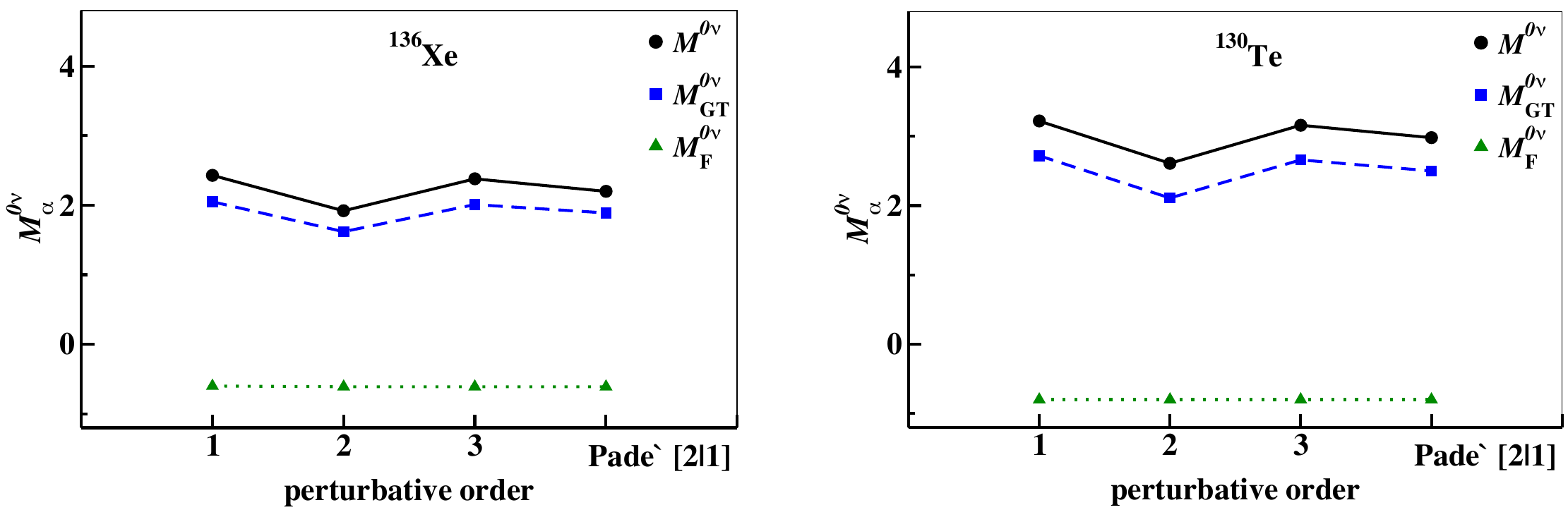}
\caption{\label{130Te136Xe_obo} \nme~for the $^{130}\mbox{Te} \rightarrow
  ^{130}$Xe and the $^{136}\mbox{Xe} \rightarrow ^{136}$Ba decay as a function of the perturbative order. The green
  triangles correspond to $M^{0\nu}_{\rm F}$, the blue squares to
  $M^{0\nu}_{\rm GT}$, and the black dots to the full \nme. Reproduced
  from Ref. \citep{Coraggio20a}.}
\end{center}
\end{figure}

Now, in order to consider the order-by-order convergence behavior in
Fig. \ref{130Te136Xe_obo} are reported the calculated values of \nme,
$M^{0\nu}_{\rm GT}$, and $M^{0\nu}_{\rm F}$ for $^{130}$Te, and
$^{136}$Xe\zbb~decay, respectively,  at first, second-, and
third-order in perturbation theory.
We compare the order-by-order results also with their Pad\'e
approximant $[2|1]$, as an indicator of the quality of the
perturbative behavior \citep{Baker70}.

It is worth pointing out that the perturbative behavior is ruled
by the Gamow-Teller component, the Fermi matrix element $M^{0\nu}_{\rm
  F}$ being only slightly affected by the renormalization procedure.
Moreover, if the order-by-order perturbative behavior of the
effective SM \zbb-decay operator is compared with the single
$\beta$-decay one, we observe a less satisfactory perturbative
behavior for the calculation of \nme, the difference between second-
and third-order results being about $30\%$ for $^{130}$Te,$^{136}$Xe
\zbb~decays.

\vspace{0.4truecm}
\section{Summary}\label{summary}
\vspace{0.1truecm}
This paper has been devoted to a general presentation of the
perturbative approach for deriving effective shell-model operators,
namely the SM Hamiltonian and decay operators.

First, we have presented the theoretical framework, which
is essentialy based on the perturbative expansion of a vertex
function, the \qbox~for the effective Hamiltonian and the
$\hat{\Theta}$-box for effective decay operators, whose calculation is
pivotal within the Lee-Suzuki similarity transformation.
The iterative procedures to solve recursive equations that provide
effective shell-model Hamiltonians have been presented in details, as
well as tips that could be useful to calculate the Goldstone diagrams
emerging within the perturbative expansion of the above mentioned
vertex functions.

In the last section, we have shown the results of a shell-model study
carried out using only single-particle energies, two-body matrix
elements of the residual interaction, and effective decay operators
derived from a realistic nuclear potential, without any empirical
adjustment.
This is a part of a large set of investigations which aim to assess
the relevance of such an approach to the study of nuclear structure.
The versatility of shell-model calculations is grounded on the ability
to reproduce experimental results for mass regions ranging from light
nuclei - $^4$He core \citep{Fukui18,Coraggio12a} - up to heavy mass
systems - nuclei around $^{132}$Sn \citep{Coraggio09d} -, as well as
to describe exotic and rare phenomena such as the Borromean structure
\citep{Coraggio10a}, quadrupole collectivity
\citep{Coraggio15a,Coraggio16a}, or the double-$\beta$ decay process
\citep{Coraggio17a,Coraggio19a,Coraggio20a} without resorting to
empirical adjustments to data.

This testifies the liveliness of this theoretical tool, and could be
inspiring for further fruitful investigations in a near and
long-distance future.

\vspace{0.4truecm}
\bibliographystyle{frontiersinHLTH&FPHY}
\bibliography{biblio}

\end{document}